\documentclass[useAMS,usenatbib]{emulateapj}

\usepackage{comment}
\usepackage{xspace}
\usepackage{graphicx}
\usepackage{epsfig}
\usepackage[backref,breaklinks,colorlinks,citecolor=blue]{hyperref}
\usepackage{appendix}

\newcommand{\kms}{km s$^{-1}$}

\def\cygx1{Cygnus~X$-$1}
\def\gx{GX~339$-$4}

\def\gro{GRO~J1655$-$40}

\def\msun{$M_{\odot}$}

\def\kms{km~s$^{-1}$}

\def\wm2{W~m$^{-2}$}

\def\cm2{cm$^{-2}$}
\def\se1{s$^{-1}$}

\def\Ave{A_{\rm V}}

\shorttitle{Compton Thick Wind in GRO J1655--40?}
\shortauthors{Neilsen et al.}

\begin{document}

\title{A Super-Eddington, Compton-Thick Wind in GRO J1655--40?}

\author{J.\ Neilsen\altaffilmark{1,2}, F.\ Rahoui\altaffilmark{3,4}, J.\ Homan\altaffilmark{1}, M. Buxton\altaffilmark{5}}
\altaffiltext{1}{MIT Kavli Institute for Astrophysics and Space Research, Cambridge, MA 02139, USA; jneilsen@space.mit.edu}
\altaffiltext{2}{Hubble Postdoctoral Fellow}
\altaffiltext{3}{European Southern Observatory, Karl Schwarzschild-Strasse 2, D-85748 Garching bei Munchen, Germany}
\altaffiltext{4}{Department of Astronomy, Harvard University, 60 Garden street, Cambridge, MA 02138, USA}
\altaffiltext{5}{Department of Astronomy, Yale University, P.O. Box 208101, New Haven, CT 06520-8101, USA}

\begin{abstract}
During its 2005 outburst, GRO J1655-40 was observed at high spectral resolution with the Chandra HETGS, revealing a spectrum rich with blueshifted absorption lines indicative of an accretion disk wind -- apparently too hot, too dense, and too close to the black hole to be driven by radiation pressure or thermal pressure (\citeauthor{M06a}). But this exotic wind represents just one piece of the puzzle in this outburst, as its presence coincides with an extremely soft and curved X-ray continuum spectrum, remarkable X-ray variability (\citeauthor{Uttley15}), and a bright, unexpected optical/infrared blackbody component that varies on the orbital period. Focusing on the X-ray continuum and the optical/infrared/UV spectral energy distribution, we argue that the unusual features of this ``hypersoft state" are natural consequences of a super-Eddington Compton-thick wind from the disk: the optical/infrared blackbody represents the cool photosphere of a dense, extended outflow, while the X-ray emission is explained as Compton scattering by the relatively cool, optically thick wind. This wind obscures the intrinsic luminosity of the inner disk, which we suggest may have been at or above the Eddington limit.
\end{abstract}


\section{INTRODUCTION}\label{sec:intro}

In outburst, accreting stellar-mass black holes exhibit an extraordinary variety of timing and spectral behaviors, commonly referred to as ``states" (see \citealt{Homan05b,RM06} and references therein). While these states are readily distinguished by their X-ray properties alone, much of what we know about them comes from simultaneous or contemporaneous radio observations (so much so that their progression in black hole outbursts is often referred to as the ``disk-jet connection" or ``disk-jet coupling;" \citealt{FBG04,Fender09} and references therein). 

Briefly, in the canonical picture, black hole transients emerge from quiescence in spectrally hard states with flat or inverted radio spectra (i.e., indicative of self-absorbed compact jets; \citealt{Corbel00,Fender01b,FBG04,Fender09}). Here, the synchrotron break is usually found between the mid-infrared and optical (\citealt{Rahoui11,Gandhi11,Corbel02,Hynes03,Buxton04,Homan05,Russell06,Coriat09}). In this state, the accretion flow is generally understood to be radiatively inefficient (\citealt{E97}), and the X-ray emission may include optically thin synchrotron emission from the jet (e.g., \citealt{Markoff01a,Homan05,Russell10}) or emission Compton scattered by a corona of hot electrons, which may be the base of the compact jet (\citealt{Markoff05}).

For a period of days to weeks, the radio and X-ray luminosities ($L_{\rm R},~L_{\rm X}$) rise in tandem ($L_{\rm R}\sim L_{\rm X}^{0.6}$, a.k.a. the radio/X-ray correlation, which is actually the low-mass end of the Fundamental Plane of black hole activity; \citealt{Hannikainen98,Corbel00,Corbel03,Gallo03,Merloni03,Falcke04,Kording06,Gallo06,Gultekin09}). The optical and infrared emission shows a strong, similar correlation, confirming its association with the compact jet (\citealt{Russell06}). Due to the increasing number of outliers, the exact nature of the correlation between $L_{\rm R}$ and $L_{\rm X}$ has been a subject of intense effort in the last several years; it is now clear that there are not one but at least two tracks in $L_{\rm R}-L_{\rm X}$ space, with some sources able to switch tracks within and between outbursts (\citealt{Coriat11,Gallo12,Corbel13} and references therein).

Eventually, on reaching a few percent of the Eddington luminosity $L_{\rm Edd}$ (e.g., \citealt{Maccarone03}), the systems undergo a state transition in which the X-ray spectral hardness and rms variability decrease significantly (\citealt{Homan05b,RM06,MunozDarias11,Motta12} and references therein), and the synchrotron emission from the compact jet is deeply quenched after a major optically-thin radio flare (\citealt{Tananbaum72,Gallo03,Homan05,Russell07,Fender09}). During the softer states, the X-ray emission is typically dominated by a multitemperature blackbody from a thin accretion disk (\citealt{SS73,Mitsuda84,McClintockRemillard06,RM06,Belloni10a,Steiner09b}). After some time the spectrum hardens and the jet reactivates before the return to quiescence (\citealt{Fender99,Corbel00,Corbel02,Gallo03,Fender09}).

\setcounter{footnote}{0}
\begin{figure*}
\centerline{\includegraphics[width=\textwidth]{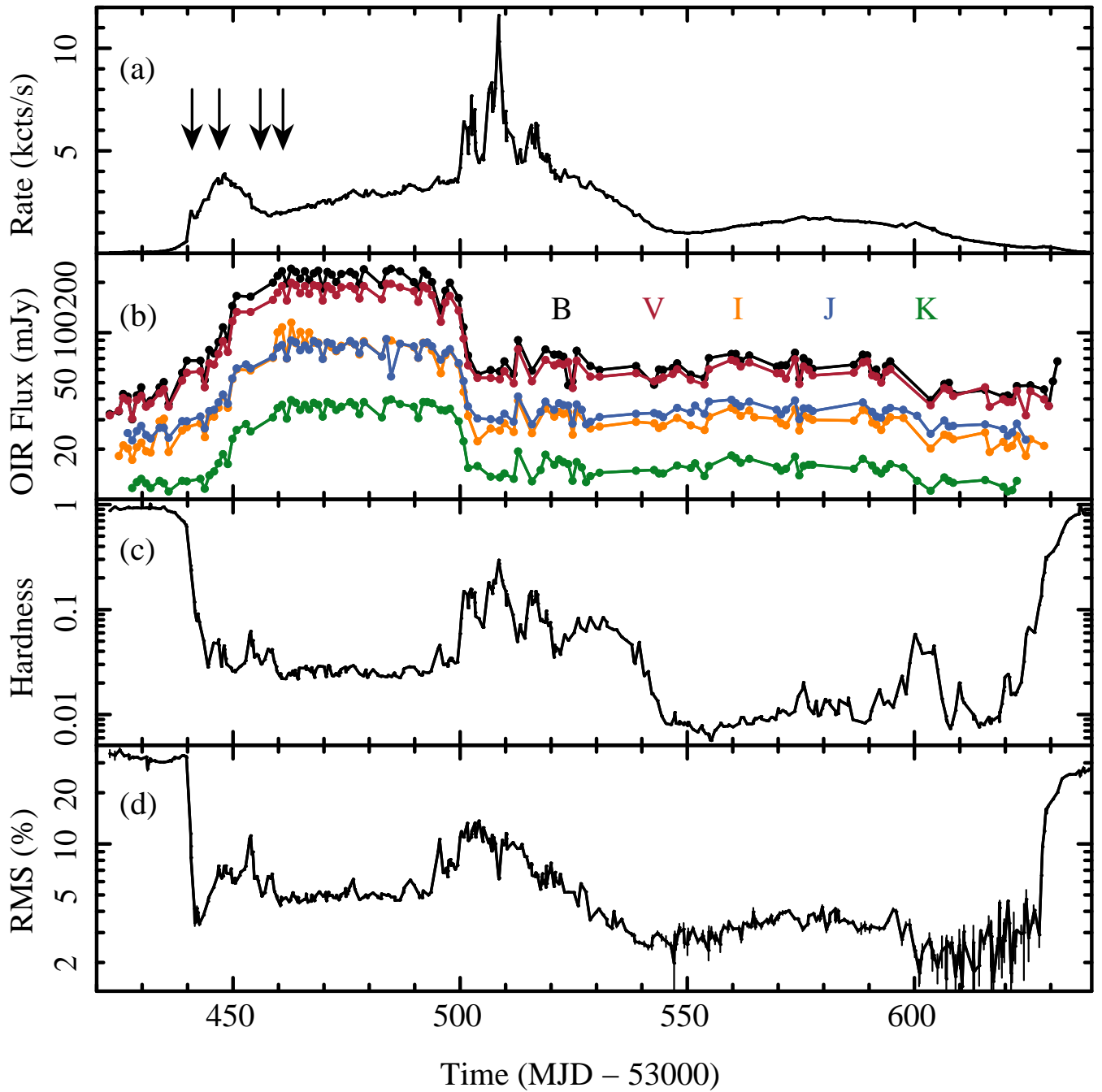}}
\caption{X-ray, optical, and infrared monitoring of GRO J1655--40 during its 2005 outburst. Panel (a): \textit{RXTE} PCA count rate (3--20 keV), with one data point per observation. Panel (b): Dereddened $BVIJK$ light curves. Panel (c): \textit{RXTE} PCA spectral hardness, defined as the ratio of the count rates in the 10--20 keV band to the 3--6 keV band. Panel (d): \textit{RXTE} PCA rms variability ($128^{-1}-64$ Hz).\label{fig:lc}}
\end{figure*}

Yet it appears that even the detailed characterizations of the evolution of black hole variability, spectral energy distributions (SEDs), and jets described above do not capture the entirety of accretion and ejection processes around stellar-mass black holes in outburst. Perhaps the most significant of recent additions to the unified picture is the accretion disk wind, a highly-ionized outflow that may be launched from the disk by an combination of radiation pressure, thermal pressure, and magnetic processes (e.g., \citealt{BlandfordPayne,B83,Woods96,Proga2000,PK02,Proga03}). In the last 20 years, a number of such absorbers have been detected around stellar-mass black holes (\citealt{Ebisawa97a,L02,M06a,M08,NL09,U09,N11a,N12a,N12b,King12,Ponti12,N13a,King14,N14a,DiazTrigo14}; see \citealt{DiazTrigo13a} for a discussion of similar absorbers in neutron star systems). These outflows are significant for two reasons in particular: (1) they can carry away the vast majority of the infalling gas (up to $\sim95\%$; \citealt{N11a,Ponti12}), thereby suppressing jets (\citealt{NL09}) and generally controlling the black hole mass accretion rate, and (2) they are found almost universally during spectrally soft states at moderate to high luminosity (\citealt{Ponti12}). These facts leave open the possibility that ionized winds play a major role in determining the outburst phenomenology of stellar-mass black holes (see also \citealt{King13}).

There are two notable exceptions to the absence of winds outside spectrally soft states. The first, GRS 1915+105 (\citealt{L02}) is exceptional in so many categories of black hole behavior that we will not dwell on it further. The second is GRO J1655-40, which was observed twice at high spectral resolution with the \textit{Chandra} High-Energy Transmission Grating Spectrometer (HETGS; \citealt{C05}) during its 2005 outburst. The first of these observations took place toward the end of a spectrally hard state and showed a single Fe\,{\sc xxvi} absorption line at 7 keV (\citealt{M08,N12b}). The second observation, 20 days later, revealed a multitude of absorption lines from O to Ni, indicative of an extremely dense, highly-ionized wind that could not reasonably be launched by thermal or radiation pressure (leaving magnetic processes as the only plausible explanation; \citealt{M06a,Netzer06,M08,Kallman09,Luketic10}). Little is known about the exact origin of the wind or what causes some winds to be launched by Compton heating and others by magnetic fields, but to date, the richness of this spectrum and the extreme properties (i.e., density, column density) of the wind are unrivaled in stellar-mass black holes.

In this paper, we report on the discovery of another exceptional property of the dense, magnetically-driven wind in GRO J1655--40: it appears to be associated with an optical/infrared (OIR) excess that is consistent with a $\sim7000$ K blackbody emitter. During the 2005 outburst, \citet{Buxton05} reported from SMARTS observations that the source remained steady in the OIR for at least 40 days. Here, we argue that this bright OIR plateau (Figure \ref{fig:lc}) cannot be explained either in the context of the disk-jet coupling described above or by any phenomena related to the companion star, and that it implies that the wind is actually Compton thick. If this OIR excess can indeed be associated with the wind, it provides a highly promising diagnostic for the presence of such winds for future observational campaigns. In Section \ref{sec:obs} we describe the observations and data reduction. In Section \ref{sec:lc} we present the multiwavelength light curves of GRO J1655--40, and we perform some modeling of the OIR SED in Section \ref{sec:sed}. We discuss and summarize our results in Section \ref{sec:discuss}.

\section{OBSERVATIONS AND DATA REDUCTION}
\label{sec:obs}

In the following, we adopt a black hole mass $M_{\rm BH}=6.3\pm0.5$ \msun, a primary to secondary mass ratio $Q=2.6\pm0.3$, an orbital period $P=2.62168\pm0.00014$~days, and an inclination $i=70.2\pm1.9$ \citep{vanderHooft98,Greene01}. The distance to the source is controversial: it was reported to be $D_{\rm BH}=3.2\pm0.2$~kpc \citep{Hjellming95}, but also argued to be lower than 1.7~kpc \citep{Foellmi06}. The ISM extinction along the line-of-sight, i.e. $\Ave=3.6\pm0.3$, is consistent with a distance in the range 3.2~kpc and 3.9~kpc \citep[see the 3D modeling of the Galactic interstellar extinction,][]{Marshall06}, and we adopt 3.2 kpc for the remainder of this paper.

\subsection{RXTE}
As detailed in \citet{M08} and elsewhere, \textit{RXTE} made frequent pointed observations of GRO J1655--40 with the PCA during its 2005 outburst. For the purposes of this paper, we focus on the observations taken between $\sim2$ weeks before and $\sim2$ weeks after the OIR plateau, covering the appearance, evolution, and disappearance of the unusual emission state. As in \citet{N12b}, we extract PCA spectra from the top layer of PCU 2, which has the best calibration, using a 0.6\% systematic uncertainty in each spectral bin. Although we extract spectra from the 3--45 keV energy band, we require a S/N ratio of at least three in each bin, which effectively limits the softest spectra to below $20-25$ keV.

\subsection{SMARTS}
As reported by \citet{Buxton05a,Buxton05}, \gro\ was observed frequently throughout the 2005 outburst with ANDICAM (\citealt{Depoy03}) on CTIO's 1.3 m telescope. Observations were made in $B$, $V$, $I$, $J$, and $K$, and the details of the data reduction are presented in \citet{Migliari07}.

\subsection{SWIFT}

Between March 6 2005 and October 25 2005, \gro\ was observed at several epochs with the {\it Swift} \citep{Gehrels04} satellite. The dataset used in this study, which is already reported in \citet{Brocksopp06}, consists of 20 observations with the Ultra-Violet/Optical Telescope \citep[UVOT,][]{Roming05} in at least one of the following filters, {\it uvw2}, {\it uvm2}, {\it uvw1}, {\it u}, {\it b}, and {\it v}. Exposure times were set between 200~s and 7800~s. We re-analyzed the data and we produced an image in each available band with {\tt uvotimsum}. We then used {\tt uvotsource} to extract the source and the background counts in 5\arcsec and 15\arcsec\ circular apertures, respectively; the derived \gro\ magnitudes are listed in \autoref{uvot}.
\begin{deluxetable*}{ccccccc}
\tabletypesize{\scriptsize}
\tablecaption{UVOT Magnitudes of GRO J1655--40 During its 2005 Outburst
\label{uvot}}
\tablehead{
\colhead{MJD}  &
{\it uvw2}&{\it um2}&{\it uw1}&{\it u}&{\it b}&{\it v}
}
\startdata
53435.5&$-$&$-$&$-$&$19.00\pm0.29$&$18.33\pm0.14$&$16.98\pm0.08$\\
53448.1&$20.04\pm0.29$&$-$&$18.53\pm0.27$&$-$&$-$&$16.07\pm0.05$\\
53449.2&$19.99\pm0.29$&$-$&$18.38\pm0.48$&$-$&$-$&$15.72\pm0.04$\\
53456.5&$19.45\pm0.12$&$20.27\pm0.58$&$17.76\pm0.14$&$-$&$-$&$15.37\pm0.03$\\
53470.5&$20.24\pm0.37$&$20.86\pm0.85$&$17.86\pm0.10$&$16.48\pm0.05$&$16.35\pm0.04$&$15.10\pm0.04$\\
53481.9&$19.33\pm0.31$&$-$&$-$&$-$&$-$&$-$\\
53504.4&$19.96\pm0.51$&$-$&$18.64\pm0.16$&$17.53\pm0.09$&$17.40\pm0.09$&$15.91\pm0.05$\\
53506.5&$20.59\pm0.52$&$-$&$19.28\pm0.27$&$18.15\pm0.13$&$17.87\pm0.09$&$16.36\pm0.06$\\
53511.4&$-$&$-$&$19.29\pm0.27$&$18.19\pm0.14$&$17.89\pm0.11$&$16.46\pm0.07$\\
53512.4&$20.78\pm0.56$&$-$&$19.79\pm0.62$&$17.53\pm0.28$&$17.96\pm0.43$&$16.05\pm0.06$\\
53525.2&$-$&$-$&$19.30\pm0.27$&$18.20\pm0.14$&$17.70\pm0.08$&$16.26\pm0.06$\\
53527.1&$-$&$-$&$20.15\pm0.58$&$18.56\pm0.20$&$18.22\pm0.12$&$16.63\pm0.08$\\
53540.1&$20.48\pm0.48$&$20.47\pm0.62$&$20.10\pm0.54$&$19.44\pm0.38$&$18.36\pm0.12$&$16.81\pm0.08$\\
53544.8&$-$&$-$&$-$&$-$&$-$&$16.37\pm0.14$\\
53636.7&$-$&$-$&$-$&$18.75\pm0.33$&$18.52\pm0.25$&$17.11\pm0.16$\\
53642.2&$-$&$-$&$-$&$-$&$18.35\pm0.20$&$17.13\pm0.12$\\
53668.5&$-$&$-$&$-$&$-$&$18.68\pm0.05$&$17.14\pm0.03$
\enddata
\end{deluxetable*}

\section{LIGHTCURVES}
\label{sec:lc}

The full X-ray/OIR monitoring of the 2005 outburst is shown in Figure \ref{fig:lc} (the X-ray outburst profile was previously presented by \citealt{M08}). As expected in the canonical picture of black hole outbursts (e.g., \citealt{FBG04,RM06,Fender09,Belloni10a}), GRO J1655-40 rose in luminosity in a spectrally hard state. A sharp transition to a bright, spectrally soft state is visible around MJD 53440 in the X-ray flux, hardness, and RMS variability. Although this state does not actually produce the softest X-ray emission during the outburst, its similarity to a `hypersoft' state in Cyg X-3 led \citet{Uttley15} to apply the same label; we adopt this term for the remainder of the paper. While the count rate of the hypersoft state rises slowly over the next $\sim50$ days, its spectral hardness and fractional RMS variability remain steady.

While the X-ray emission at first glance appears fairly typical of black hole outbursts, the OIR lightcurve is unusual. In spectrally hard states, this emission is normally associated with the jet (e.g., \citealt{Russell06,Migliari07,Russell10,Rahoui11} and references therein) and is expected to be quenched along with the jet during the transition to a brighter, softer state (\citealt{Homan05}). But in this outburst, the OIR emission rises after the transition to the hypersoft state, reaching as much as $5-10$ times its pre-transition level. This excess is clearly visible in the 2005 April 6 spectral energy distribution (SED) shown by \citet{Migliari07}, who noted that it was consistent with a blackbody but focused on other portions of the outburst. As with the integrated timing properties noted in the preceding paragraph (see also \citealt{Uttley15}), the OIR excess emission is generally steady throughout the hypersoft state; it also appears to be uncorrelated with the X-ray flux. 
\begin{figure}
\centerline{\includegraphics[width=0.48\textwidth]{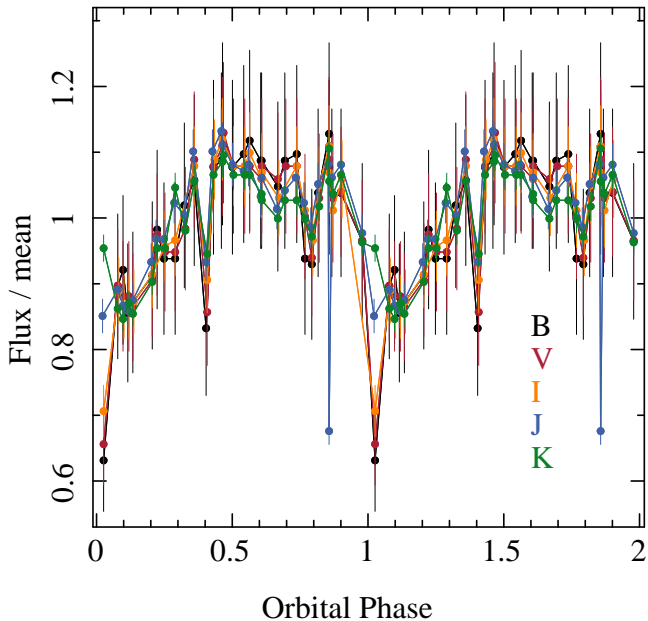}}
\caption{Phase-folded, normalized light curves of the bright OIR plateau in the 2005 outburst of GRO J1655--40. Two cycles are shown for clarity, and we have excised the four brightest $I$-band points (see Figure \ref{fig:lc}). The orbital modulation has a fairly smooth, somewhat sinusoidal profile that is similar at all wavelengths. There is a deep minimum at phase $\sim0$ that is deeper at shorter wavelengths.\label{fig:phaselc}}
\end{figure}

\subsection{Orbital Variability}
In the course of inspecting the OIR plateau, we noticed that the scatter in the $BVIJK$ light curves appears to be potentially periodic, reminiscent of a steady flux level with recurrent dips. Indeed, the plateau fluxes (here defined as MJD 53459 -- 53499) exhibit an oscillation with a fractional rms variability of $\sim10\%$ at all wavelengths when folded on the orbital period of \citet{vanderHooft98}\footnote{This is defined such that inferior conjunction, the closest approach between observer and donor star, occurs at phase 0.}. The oscillation is shown in the normalized, phase-folded $BVIJK$ light curves in Figure \ref{fig:phaselc}; two cycles are shown for clarity. The orbital profile is very similar at all wavelengths: a smooth oscillation with a broad maximum between 0.5 and 0.9, with a minimum around phase 0.1 and a few dips that are deeper at shorter wavelengths (and may therefore be unrelated to the smooth oscillation). Due to the large number of orbits completed since the \citet{vanderHooft98} reference point ($\sim1,395$), the phases quoted here are sensitive to the orbital period, with a systematic uncertainty of roughly 0.1-0.2 in phase. To determine robust phases, it may be necessary to use the SMARTS data to refine the orbital period.

To verify the statistical significance of this oscillation, we performed an epoch-folding analysis (\citealt{Davies90}) using S-Lang/ISIS Timing Analysis Routines\footnote{http://space.mit.edu/CXC/analysis/SITAR/functions.html} (SITAR). First, we consider each wavelength independently, folding the plateau light curve using 50 evenly spaced trial periods from 1 to 10 days and 5 phase bins in the folded light curve. The resulting $L$ statistics show a noticeable peak at roughly the orbital period $P_{\rm orb}$, as well as at $2P_{\rm orb}$ and $3P_{\rm orb},$ lending credence to the idea that this is truly an oscillation related to the orbital period of GRO J1655-40.

\begin{deluxetable*}{cccc@{\extracolsep{3mm}}ccccc}
\tabletypesize{\scriptsize}
\tablecaption{Near-IR to UV SED Parameters for GRO J1655--40 
\label{tbl:bestpar}}
\tablewidth{0pt}
\tablehead{
\colhead{}&
\multicolumn{3}{c}{{\tt blackbody + powerlaw}} &
\multicolumn{5}{c}{{\tt baseline blackbody + hypersoft blackbody excess + powerlaw}}\\
\cline{2-4}\cline{5-9}\\[-1.5mm] 
\colhead{MJD}&
\colhead{$R_{11}$}&
\colhead{$T$ (K)}&
\colhead{$F_{0.2}$}&
\colhead{$R_{\rm b,11}$}&
\colhead{$T_{\rm b}$ (K)}&
\colhead{$R_{\rm ex,11}$}&
\colhead{$T_{\rm ex}$ (K)}&
\colhead{$F_{0.2}$}
}
\startdata
53448.1 & $4.7_{-0.4}^{+0.5}$ & $5000\pm500$ & $72\pm13$ & $4.5\pm0.2$ & $4100\pm200$ & $2.3\pm0.1$ & $7200\pm400$ & $60\pm11$\\
53449.2 & $5.0_{-0.4}^{+0.4}$ & $5500\pm500$ & $96_{-21}^{+20}$ & 4.5\tablenotemark{a} & 4100\tablenotemark{a} & $3.1\pm0.2$ & 7200\tablenotemark{c} & $79_{-16}^{+16}$\\
53470.5 & $6.3_{-0.3}^{+0.3}$ & $6500\pm300$ & $100_{-18}^{+17}$& 4.5\tablenotemark{a} & 4100\tablenotemark{a} &  $5.2\pm0.2$&  7200\tablenotemark{c} & $91\pm18$\\
53506.5 & $4.1_{-0.4}^{+0.5}$ & $4500\pm500$ & $31\pm5$ & 4.5\tablenotemark{a} & 4100\tablenotemark{a}& \nodata&  \nodata & $35\pm4$\\
53511.4 & $4.3\pm0.5$ & $4100_{-400}^{+500}$ & $28_{-5}^{+4}$& 4.5\tablenotemark{a} & 4100\tablenotemark{a} &\nodata&  \nodata&  $29\pm4$\\
53512.4 & $5.5_{-0.5}^{+0.6}$ & $4400\pm400$ & $65\pm11$ & $6.0\pm0.3$\tablenotemark{b} & 4100\tablenotemark{a} & \nodata & \nodata & $71\pm9$\\
53525.2 & $4.6\pm0.5$ & $3900_{-300}^{+400}$ & $24\pm4$ & 4.5\tablenotemark{a} & 4100\tablenotemark{a} & \nodata&  \nodata & $22\pm4$
\enddata
\tablecomments{Best-fit to the extinction-corrected UVOT+SMARTS SEDs of \gro\ during its 2005 outburst. The model is the combination of (1) the expected stellar emission from the F6IV companion star; (2) a UV power law with a spectral index 1.6; and (3) spherical blackbodies to account for the excess OIR emission. For the parameters on the left side of the table, this is the complete model. For the parameters on the right, we separate the OIR excess into one steady baseline component and one excess during the hypersoft state. $R_{11},$ $R_{\rm b,11},$ and $R_{\rm ex,11}$ are the radii of the blackbody, the steady baseline blackbody, and the hypersoft excess blackbody in units of $10^{11}$ cm; the $T$ columns are the corresponding blackbody temperatures. $F_{0.2}$ is the power law flux density at 0.2 $\mu$m in mJy. Uncertainties are given at $1\sigma$.}
\tablenotetext{a}{Tied throughout the outburst.}
\tablenotetext{b}{Flares on MJD 53512 required us to vary the blackbody radius for this SED.}
\tablenotetext{c}{The temperature of the blackbody excess is constrained to be constant during the hypersoft state.}
\end{deluxetable*}

Complicating this interpretation are the observational errors, which are not accounted for by the epoch folding routine: while the oscillation is most significant in $B$ and $V$ band light curves (with implied single-trial significance levels of $\gtrsim10^{-5}$), these bands also have the largest observational errors. But while we may not have the statistics to measure the oscillation period independently from the plateau data, the $L$ statistic is ideal for confirming periodicities (\citealt{Davies90}), so we can focus on epoch folding at the known orbital period. In order to incorporate the sizable observational errors, we exploit the similarity of the oscillation profile at all wavelengths using a weighted mean of the normalized fluxes from each day during the plateau. More specifically, we divide each of the $BVIJK$ time series by its respective mean and then compute daily weighted averages and standard deviations, with weights given by the inverse squared error on the daily fluxes. For 10,000 Monte Carlo iterations, we sample the observational errors and calculate the $L$ statistic at the orbital period. The average $L$ statistic corresponds to a detection at 99.68\% confidence, i.e., just below $3\sigma$. For slightly different periods (e.g., 2.62191~d as determined by \citealt{Greene01}) we find significance levels just above $3\sigma.$ Thus it appears that the OIR plateau is modulated on (or near) the orbital period with an rms amplitude of about 10\%. No  modulation is detectable in the X-ray emission.

\section{Spectroscopy}
\label{sec:sed}

\subsection{Spectral energy distributions}

To understand the origin of the low-energy emission throughout the hypersoft state, we model the \gro\ UV to near-IR SEDs. We built them using UVOT and SMARTS photometry, based on the following three criteria: (1)  the observations took place in the soft or hypersoft states, which rules out any significant contribution from the compact jets; (2) the UVOT and SMARTS data were quasi-simultaneous, i.e. within a few hours of each other; and (3) \gro\ was detected in the {\it uw2} and/or {\it uw1} filters to better constrain any emission in the UV. Seven UV to near-IR SEDs built with data obtained between MJD~53448 and MJD~53527 fulfilled these conditions, including one during the plateau phase on MJD~53470. We corrected them for interstellar absorption $E(B-V)\,=\,1.2\pm0.1$ \citep{Hynes98} using the extinction law given in \citet{Fitzpatrick99} for a total-to-selective extinction ratio $R_{\rm V}=3$, the expected value along \gro\ line-of-sight \citep{Wegner03}. 

\subsubsection{Modeling}

We first consider a combination of stellar and accretion disk emission. \gro's companion star was first identified as an F3-6IV sub-giant \citep{Orosz97}. This identification was later confirmed and refined to F6IV \citep{Foellmi06}, and a significant stellar contribution to the optical and near-IR domain is therefore expected. Indeed, the B and V magnitudes of \gro\ in quiescence are clustered around 18.7 and 17.2, respectively \citep{Greene01}, which is consistent with our own measurements with UVOT at the end of the 2005 outburst (see \autoref{uvot}). This is still relatively significant compared to the minimum B and V magnitudes measured during the plateau phase, i.e. around 16.4 and 15.0, respectively, and we therefore model the UV to near-IR emission of the companion star by scaling an F6IV Kurucz synthetic spectrum to the extinction-corrected flux densities in quiescence. X-ray reprocessing from the inner to outer regions of the accretion disk is thought to dominate the UV emission of microquasars in the soft state (e.g., GRS 1915+105, \citealt{Rahoui10}; \gx, \citealt{Rahoui12}), so we add a power law $F_{\nu}\propto\nu^{1.6}$, consistent with irradiation. Allowing the power law index to vary leads to flatter slopes but requires no significant changes to our major conclusions. Note that we did not try to use more appropriate irradiated accretion disk models such as {\sc diskir} \citep{Gierlinski09} -- which would have allowed us to fit both high-energy and low-energy data -- because, as shown below, no direct disk emission is detected in the X-ray band during the hypersoft state.

\begin{figure*}
\centerline{\includegraphics[width=\textwidth]{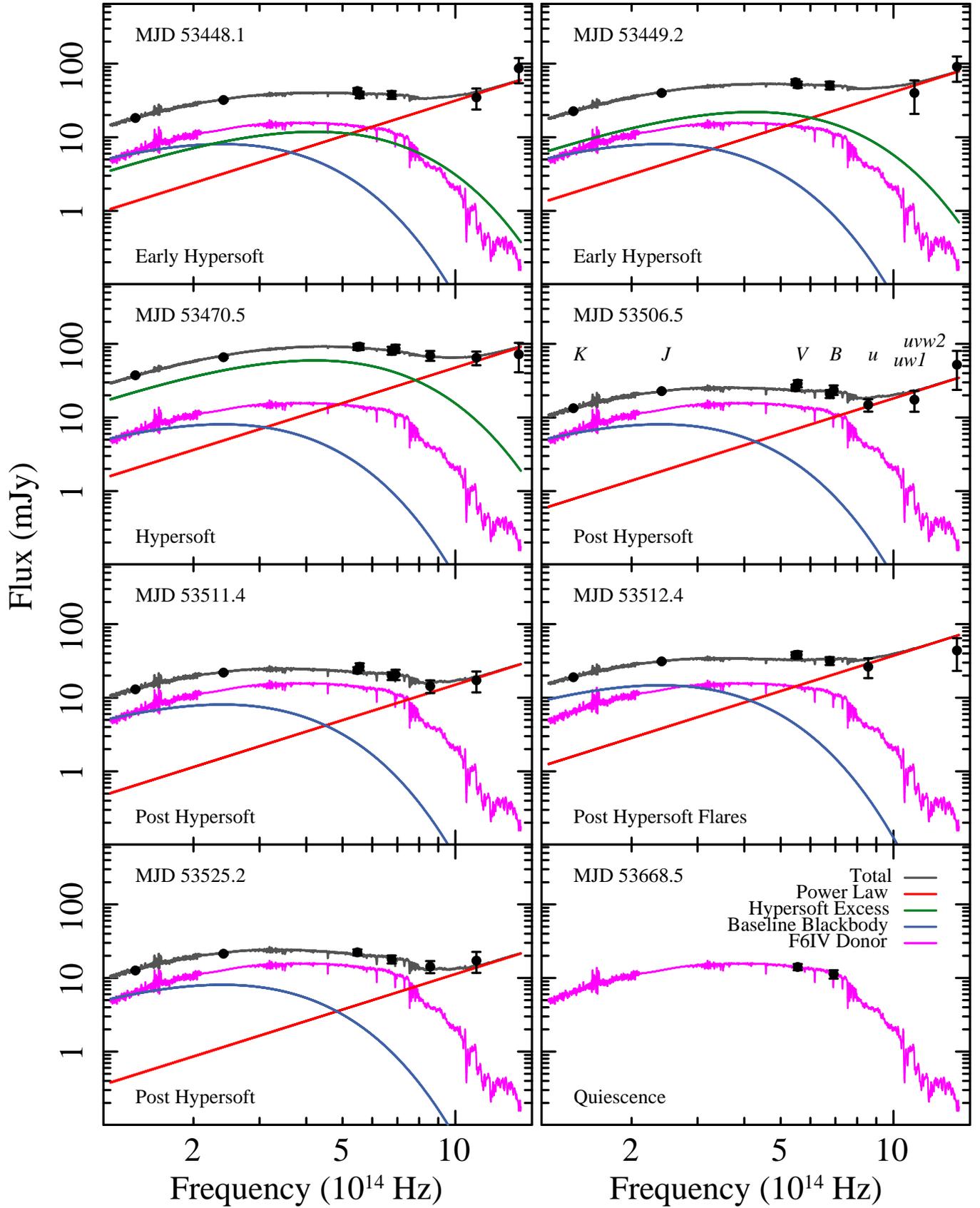}}
\caption{\small Best-fit to the extinction-corrected UVOT+SMARTS SEDs of \gro\ during its 2005 outburst. The model is the combination of (1) the expected stellar emission from the F6IV companion star; (2) a UV power law with a spectral index 1.6; (3) a spherical blackbody to account for the steady OIR baseline emission; and (4) a spherical blackbody to represent the excess emission during the hypersoft state.\label{fig:bestsed}}
\end{figure*}

Due to the presence of an optical and near-infrared excess in all SEDs, the combination of the stellar continuum and the irradiated disk results in unsatisfactory fits ($\chi^{2}_\nu=5020.12/46=109.11$ over all datasets). To account for this excess, we add a spherical blackbody component, which dramatically improves the fits ($\chi^{2}_\nu=36.14/32=1.14$). But because there is a long period (MJD 53500 - MJD 53600) with a steady optical/infrared excess unrelated to the hypersoft state, we also consider a model that includes both this baseline blackbody and an additional blackbody during the hypersoft state (see Figure \ref{fig:bestsed} and Table \ref{tbl:bestpar} for the best-fit SEDs and parameters, respectively). While two blackbodies are not statistically required during the hypersoft state, this model is motivated by the physical behavior of the system (two apparent emission components in the lightcurve), and it provides a much better fit for fewer free parameters ($\chi^{2}_\nu=33.25/39=0.85$). 

\subsubsection{The optical and near-infrared excess}
\label{sec:xs}
For our initial blackbody/power law model for the IR to UV SED of GRO J1655-40, we find blackbody temperatures ranging from 3900 K to 6500 K and radii ranging from $4.1\times10^{11}$ cm to $6.3\times10^{11}$ cm; the power law flux dominates the UV emission (see Figure \ref{fig:bestsed}) at a level of 24-100 mJy. Interestingly, the blackbody radius and temperature in this model are strongly correlated with the power law flux (correlation coefficients of 0.8 and 0.9, respectively). The interpretation of this correlation depends on the origin of the power law -- we have fixed the power law index at 1.6 to simulate irradiation, but in reality this is poorly constrained by our small number of data points. 

The parameters in our preferred model are fairly similar. The baseline blackbody has a radius $R_{\rm b}=4.5\pm0.2\times10^{11}$ cm and a temperature $T_{\rm b}=4100\pm200$ K, consistent with the results presented in the previous paragraph. Meanwhile, the radius of the blackbody excess $R_{\rm ex}$ rises from $(2.3\pm0.1)\times10^{11}$ cm on MJD 53448.1 to $(3.1\pm0.2)\times10^{11}$ cm on MJD 53449.2 to $(5.2\pm0.2)\times10^{11}$ cm on MJD 53470.5, during the hypersoft state proper. This hypersoft excess has a temperature of $T_{\rm ex}=7200\pm400$ K.

One complication is the observation on MJD 53512.4, which appears to be affected by the bright flares after the hypersoft state. For this observation, if we let the baseline blackbody radius vary, it increases to $6.0\pm0.3\times10^{11}$ cm. If we instead allow a brief return of the hypersoft blackbody with $T=7200$ K, the radius is $(2.5\pm0.1)\times10^{11}$ cm.

Given their large sizes, it is worth comparing the measured blackbody radii to characteristic orbital size scales in GRO J1655--40. For the orbital parameters adopted above, the effective Roche lobe radius and tidal radius of the accretion disk are $R_{\rm L}\sim0.47a$ and $R_{\rm tide}\sim0.42a$, respectively, where $a=1.1\times10^{12}$ cm is the semi-major axis \citep[see formulas in][]{Eggleton83,FKR02,Leahy15}. The baseline blackbody has a radius $\sim70\%$ larger than the Roche lobe of the secondary star, and therefore may include contributions not only from the heated face of the companion but also the outer accretion disk and/or hot spot. The excess is more interesting. Because the outer radius of the accretion disk $R_{\rm out}$ should not be larger than $R_{\rm tide}$, we find $R_{\rm out} \le 4.8\times10^{11}$ cm, meaning that at maximum, i.e., during the plateau/hypersoft state, the component responsible for the blackbody excess enshrouds the whole accretion disk.

\subsection{X-ray Spectra of the Hypersoft State}
\label{sec:spec}
To support characterizing the X-ray data from MJD 53450--53500 as a hypersoft state, \citet{Uttley15} pointed to the unusual shape of the X-ray spectrum, which is not well described by the standard model of a disk blackbody and a Comptonized power law. Indeed, for three sample spectra, a joint fit of an absorbed, Comptonized disk blackbody (\citealt{Steiner09b} and references therein) plus an additional absorption edge resulted in a reduced $\chi^2$ in excess of 10. We noted in \citet{N12b} that modeling the hard component with {\tt nthcomp} (\citealt{Zdziarski96,Zycki99}) or including a high-energy cutoff can lead to satisfactory spectral fits during this period. Thus, alternatives to the standard disk-plus-power law model are desirable. In this section, we briefly explore the behavior of this standard model, which we call Model 1, and then consider two alternative continuum models: {\tt ezdiskbb+nthcomp} and {\tt ezdiskbb+bremss} (Models 2 and 3, in which we separately replace the power law with {\tt nthcomp} and bremsstrahlung, respectively). 

In addition to the continuum components described above, each fit includes a Gaussian emission line in the 5-8 keV range, a smeared absorption edge to account for the effect of ionized absorption (both lines and edges; {\tt smedge}), and interstellar absorption ({\tt TBnew}; \citealt{Wilms00}). We fix $N_{\rm H}<10^{22}$ cm$^{-2}$ based on the 0.5--8 keV \textit{Chandra} Medium Energy Grating (MEG) spectrum first published by \citet{M06a}. The continuum components are normalized by their 3-25 keV unabsorbed fluxes using the {\tt cflux} convolution model. We perform these spectral fits within the Interactive Spectral Interpretation System (ISIS; \citealt{HD00,Houck02}).

\subsubsection{Model 1: ezdiskbb+powerlaw}
\label{sec:model1}
Since \citet{Uttley15} already noted that Model 1 provides formally unacceptable fits to the X-ray spectra of the hypersoft state, we shall not dwell on it in great detail (for clarity, we omit the evolution of its parameters from Figure \ref{fig:spec}). It is useful, however, to the extent that it provides a sense of what to expect from our alternative models in subsequent sections. 

In Model 1, the pre-hypersoft portion of the outburst proceeds normally, beginning in a state dominated by a hard power law with a photon index $\Gamma\sim1.4-2.2.$ As the total flux rises, so does the fraction of the flux contributed by the disk; by the initial peak of the outburst around MJD 53447, the disk contributes $\sim90\%$ of the total flux. After this, however, the results begin to deviate from typical black hole behavior. The disk component begins to fade and the power law simultaneously brightens and softens (reaching our upper limit of $\Gamma\sim5$ within a few days). In addition, the smedge optical depth $\tau_{\rm sm}$ rises to roughly 3 by MJD 53455 and remains steady for the duration of the hypersoft state. In general, there is fairly little evolution of spectral parameters during this plateau, and the model returns a typical reduced $\chi^2$ of 1.5-3. After MJD 53500, the power law returns to normal values between 2 and 3, $\tau_{\rm sm}$ decreases to $\sim0,$ and the fits become statistically acceptable again. The upshot, which echoes the results of \citet{Uttley15}, is that the X-ray spectrum of the hypersoft state is quite steep but not well described by a power law, in part due to significant absorption structure in the 5-10 keV band, but also due to strong curvature in the broadband continuum. 

\subsubsection{Model 2: ezdiskbb+nthcomp}
\label{sec:model2}
One option for introducing curvature into the spectrum is a more physically-motivated Comptonization model (as the power law primarily represents the case of a scattering medium with high temperature and low optical depth). As noted above, for Model 2 we replace the power law in Model 1 with {\tt nthcomp}. Since we are interested in the scattering optical depth $\tau$, we constrain the photon index $\Gamma$ to be a function of $\tau$ and the electron temperature $kT_{\rm e}$ according to the formula given by \citet{Sunyaev80,U09}:
\begin{equation}
\Gamma = \sqrt{2.25+\frac{3}{(kT_{\rm e}/511~{\rm keV})((\tau+1.5)^2-2.25)}}-0.5.
\end{equation} Where possible (i.e., where $\chi_{\nu}^{2}<1.5$ without the disk component) we fit the spectrum with {\tt nthcomp alone}; otherwise, we tie the temperature of the seed photons---which have a blackbody spectrum---to the disk temperature. According to Model 2, much of the outburst proceeds as in the initial analysis. Prior to the hypersoft state, the Compton component dominates the flux, and no disk is required for the earliest phase of the outburst. Here $kT_{\rm e}$ is large (in most cases consistent with 1 MeV but generally constrained to be above 10 keV) and $\tau\lesssim0.3$, such that the Compton component mimics a power law over the observed energy range. 
\begin{figure*}
\centerline{\includegraphics[width=\textwidth]{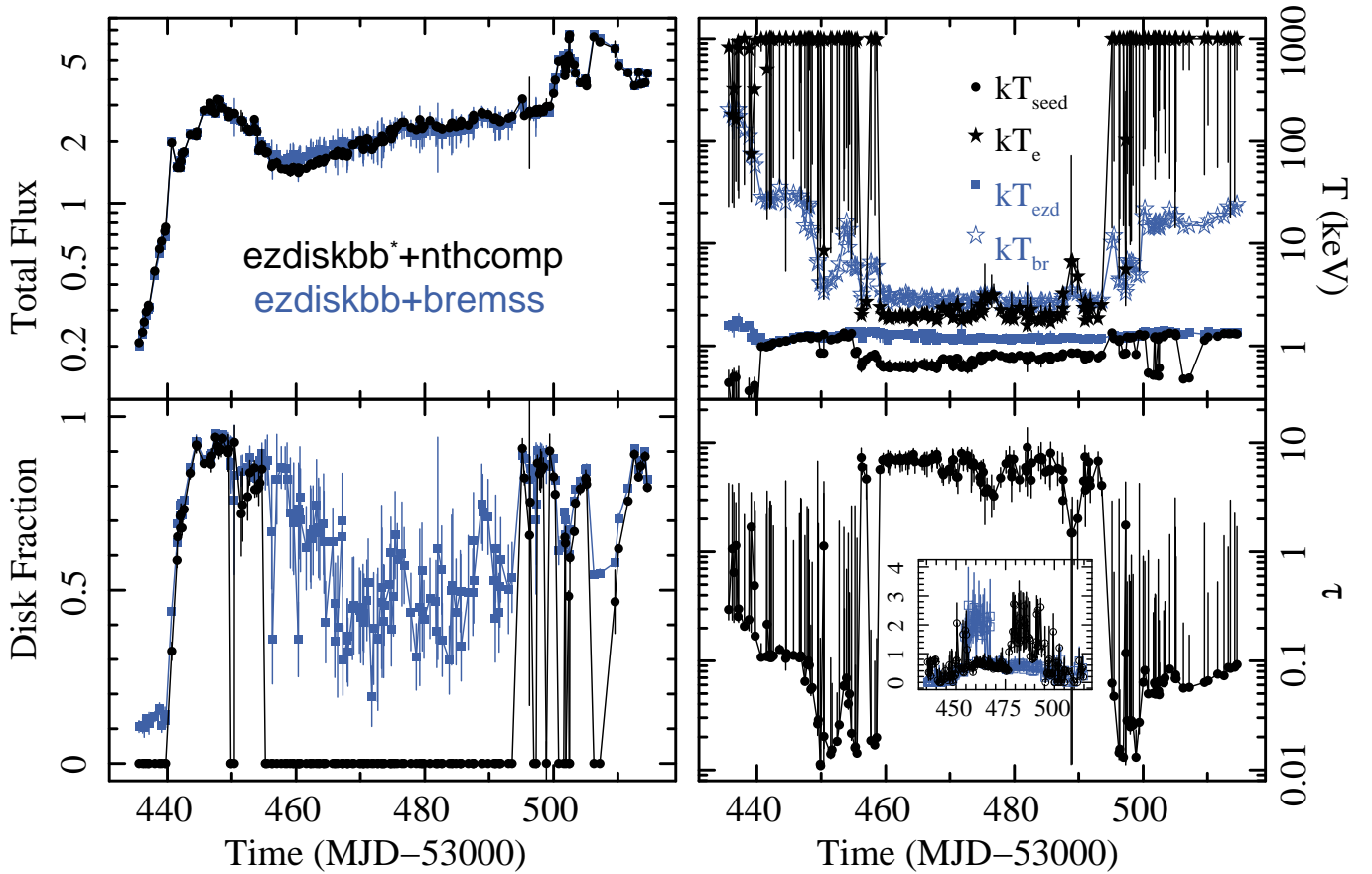}}
\caption{X-ray spectral parameters for Models 2 and 3. {\tt ezdiskbb}* in Model 2 indicates that the disk is only included when it is statistically required. The soft ratio is the fraction of the total flux emitted by the soft component ({\tt ezdiskbb} or {\tt bremss}). In the bottom right panel, $\tau$ is the optical depth in the thermal Comptonization component; the inset shows the optical depth in the smeared absorption edge. See text for details.\label{fig:spec}}
\end{figure*}

As the flux rises to its peak, the equivalent photon index $\Gamma$ rises from $\sim1.5$ to $\sim2.2$, $\tau$ decreases, and the disk comes to dominate the spectrum (see Figure \ref{fig:spec}). But then, around MJD 53450, the disk fraction begins to fall again, and within a few days there is no evidence for a direct disk component. Instead, the spectrum appears to be that of a low temperature ($kT_{\rm e}\gtrsim1.7$ keV), high optical depth ($\tau\sim5-8$) scattering medium. Although the exact evolution is difficult to track, because manually removing the disk component when it is not required enhances some of the sharp changes in the fit parameters, it is presumably this optically thick scattering medium that is responsible for the effective disappearance of the disk. During the hypersoft state, the scattering optical depth decreases slowly and the electron temperature rises slowly, until the spectrum begins to return to its pre-hypersoft configuration around MJD 53495. During the period when the electron temperature is lowest, the total fit statistic (combined over all 86 spectra) is $\chi^2/_\nu=2310.864/2573=0.898.$

\subsubsection{Model 3: bremss+powerlaw}
\label{sec:model3}
The sudden change in the behavior of the Compton component in Model 2, from a hot optically thin medium to a much cooler medium with a much higher optical depth raises some questions about the physical plausibility of this model (see Section \ref{sec:discuss}). As an alternative, we consider Model 3, in which we replace the power law emission in Model 1 with bremsstrahlung (bremsstrahlung accounts for the curvature in the spectrum at higher energies better than the power law).

While bremsstrahlung is rarely used to model X-ray emission from stellar mass black holes\footnote{Interestingly, \citet{Zdziarski10} argued that bremsstrahlung makes a significant contribution to the hard state X-ray spectrum in Cyg X-3, but to our knowledge this argument has not been applied to the hypersoft state.}, the first thing to note about the results of Model 3 is their similarity to those of Model 2: 
\begin{itemize}
\item The total flux light curve (Figure \ref{fig:spec}) is nearly indistinguishable between Models 2 and 3, and the smeared edges are also somewhat similar (with the exception of some discontinuities attributable to the combined uncertainty in the location and strength of the edge and iron line).
\item Outside the hypersoft state, the bremsstrahlung component contributes the same fraction of the total flux as the Compton component in Model 2, and the disk temperatures in Model 2 and 3 are very similar.
\item The bremsstrahlung temperature is similar to the electron temperature in the hypersoft state, and its evolution during the rest of the outburst is very similar to that of the scattering optical depth in Model 2.
\end{itemize}
In other words, Model 3 captures the same qualitative spectral evolution described in Section \ref{sec:model1}: a normal hard component with a sharply brightening soft component, followed by a hypersoft state dominated by a steep and curved component with significant absorption structure, and finally a return to typical soft/hard component behavior after the hypersoft state. The fit quality, however, is not as good: the total fit statistic is $\chi^2/_\nu=2820.558/2574=1.096.$

Where Model 3 diverges from Models 1 and 2 is in the physical characterization of the hypersoft state itself. Clearly the same spectral evolution is reflected in the fit parameters, but while one model attributes this to an optically thick scattering medium, the latter model explains it as thermal emission from an optically thin cloud. We discuss the significance of these results in Section \ref{sec:discuss}.
 
\section{DISCUSSION}
\label{sec:discuss}

A comprehensive explanation of the multiwavelength variability properties of the hypersoft state in GRO J1655--40 is beyond the scope of this work. \citet{Uttley15} have already presented models for the timing behavior of the system, and the high-resolution X-ray spectral observations of the disk wind have been analyzed in detail. Here we focus on three specific issues: (1) the nature of the unusual X-ray continuum during this state, (2) the origin of the OIR blackbody excess, and (3) the connection between these phenomena and the accretion disk wind. We shall argue that not only are these processes inextricable, but also that the wind may be responsible for both the X-ray emission and the OIR excess. In the OIR, we restrict our attention to the hypersoft excess, on the idea that the baseline blackbody component can be attributed to some combination of the irradiated face of the donor star and emission from the outer disk and/or hot spot.

For clarity, we offer a brief sketch of the structure of this section. First, in Section \ref{sec:corona}, we suppose that the observed X-ray emission is produced in the inner accretion disk and corona and subsequently absorbed by an ionized wind at $\gtrsim10^{9}$ cm from the black hole. This scenario does not provide a satisfactory interpretation of the OIR excess (or for the existence of a cool, dense corona), so we consider modifications in which the X-ray emission arises in the wind, whether due to Compton scattering (Section \ref{sec:coronawind}) or thermal emission (Section \ref{sec:wind}). We conclude that a Compton-thick wind provides the best explanation for the properties of the hypersoft state.

\subsection{Hypersoft State due to an Optically Thick Corona}
\label{sec:corona}

In Section \ref{sec:spec}, we showed that the X-ray continuum during the hypersoft state could be described as emission from an accretion disk Comptonized by cool electrons with a high optical depth (Model 2). From the perspective of the canonical behavior of black holes in outburst (e.g., \citealt{FBG04,RM06,Belloni10b}), this model is not completely intuitive, but we begin here because the Comptonized accretion disk model is at least superficially consistent with expectations.

\subsubsection{The X-ray spectrum}
The salient feature of Model 2 is a precipitous drop in the electron temperature coupled with a sharp rise in the scattering optical depth roughly 10 days after the peak of the outburst (i.e., MJD $\sim53459$). The error bars on $kT_e$ are significant, but the best fit value drops from $\sim990$ keV to 2.4 keV between MJD 53458.58 and 53459.10. These conclusions are robust to the Comptonization model used: {\tt simpl} (\citealt{Steiner09b}) modified to allow a high-energy cutoff, {\tt nthcomp}, {\tt comptt,} and {\tt eqpair} (\citealt{Coppi99}) all lead to similar conclusions. The final temperature is unusually low for coronal electrons, and is particularly atypical for spectrally soft states (where some lines of evidence indicate that non-thermal electrons may dominate Comptonization processes, e.g., \citealt{Gierlinski99,Done07} and references therein). 

The main challenge, then, for a physical scenario relating the hypersoft state to an optically thick corona around the black hole is explaining why the electrons cool so much. Without detailed constraints on the size and geometry of the emitting region, it is difficult to say much about the energetics of Model 2 and its variants. However, a cursory analysis with {\tt eqpair} (specifically, a comparison between a pre-hypersoft observation and a hypersoft observation, on MJDs 53444.49 and 53463.48, respectively) indicates that during the hypersoft state, there is marginally more power in the electron acceleration process than in the seed photons\footnote{Since this is essentially the ratio between heating and cooling, it is a major determining factor for the electron temperature (see, e.g., \citealt{Done03,Done07} and references therein). The models described here are for purely thermal Comptonization.}. The model also indicates a $\sim20\%$ decrease in the normalization (i.e., seed photon flux; \citealt{Gierlinski99,Nowak12b}) during the hypersoft state, but it appears that the sharp decrease in electron temperature is driven primarily by a $\sim100\times$ increase in the scattering optical depth.

What, then, is the origin of the increase in $\tau?$ For a fixed number of electrons, the optical depth will rise if the scattering medium becomes more compact. But it does not seem plausible for a typical corona to shrink by a factor of 100, and there is no self-evident way to understand the association between such a small corona and the large blackbody excess described in Section \ref{sec:xs}. As we argue in the next section, unrelated explanations of the OIR excess are unsatisfactory. Thus we are left with the alternative: there must be more cool (cooling) electrons along the line of sight.

\subsubsection{The optical/infrared excess}
\label{sec:star}
Interpreted as the appearance of a compact, optically thick corona, Model 2 sheds little light on the nature of the OIR excess, a $\sim7000$ K blackbody with a radius comparable at peak to the size of the accretion disk. Since the OIR emission is expected to drop in soft states even in scenarios where it is produced by a hot accretion flow or corona (e.g., \citealt{Veledina13}), here we consider whether it is possible to attribute this emission to the accretion disk or to the companion star, and whether they could produce the $\sim10\%$ variability with orbital phase. 

While it is tempting to attribute the excess to the disk itself, there is a significant problem with this interpretation: the disk area is too small. At the orbital inclination of \gro, a disk with the same projected area as the spherical blackbody would have a radius greater than the orbital separation of the binary. Furthermore, emission from the outer disk (either intrinsic or irradiated) is already included in our SED modeling via the power law and the baseline blackbody. The disk does not appear to present a viable explanation for the excess, although the $\sim10\%$ variability is comparable to what is seen in superhumps (e.g., \citealt{Smak06}).

At face value, a stellar interpretation is even more tempting, since the temperature of the blackbody during the hypersoft state is comparable to that of the F6IV companion (Figure \ref{fig:bestsed}), and the excess exhibits orbital variability. However, this interpretation encounters a similar problem to the blackbody-in-the-disk scenario: the area of the blackbody excess is a factor of $\sim2.3$ larger than the surface area of the companion star (i.e., its Roche lobe; \citealt{Leahy15}). Furthermore, as with the disk, the stellar contribution is likely accounted for by the baseline blackbody. Finally, the wavelength-independent amplitude of the smooth OIR variability is not consistent with the strong wavelength dependence of the ellipsoidal modulations of the secondary (\citealt{Greene01}). Together, these points indicate that the excess is likely unrelated to the companion.

As an aside, it is worth noting that if we use \citet{Foellmi06}'s distance to GRO J1655--40 of 1.7 kpc, the stellar surface area problem may be lessened or alleviated. The other problems we mentioned would still remain, however, in addition to the failure of the secondary to fill its Roche lobe (see the private communication in \citealt{M08}). It therefore seems likely that the OIR excess is related to origin of the X-ray properties of the hypersoft state, which further rules out the compact cold corona hypothesis for the X-ray emission.

\subsection{X-rays from the Wind?}

In Section \ref{sec:corona}, we considered Model 2, in which we treated the X-ray spectrum as disk emission Comptonized by a hot corona. Although the spectral fits were good, it was difficult to interpret the results without invoking the appearance of a new population of electrons. Fortunately, observations indicate the sudden appearance of a large column density of highly-ionized X-ray absorbing gas (the accretion disk wind), and there is nothing in the fit model specific to the corona; could the free electrons in the wind be responsible for scattering the disk emission during the hypersoft state? Or (Section \ref{sec:wind}) could the wind directly emit the X-rays?

\subsubsection{Scattering in the wind}
\label{sec:coronawind}
There are two primary complications for the scattering scenario. First, the observed electron optical depth in the wind $\tau_{\rm w}\sim0.6$ (i.e., $N_{\rm H,w}\lesssim10^{24}$ cm$^{-2}$; \citealt{Kallman09}) is less than the scattering optical depth $\tau_{\rm e}\sim5$. But the measured column density only probes a narrow region in ionization space, and there may be a significant amount of fully ionized material in the wind that is not visible in X-ray line absorption. For reference, if the average density and extent of the fully ionized portion of the  wind are comparable to the same quantities measured in the X-ray absorber ($n_{e}\sim10^{15}$ cm$^{-3}$, $R<7\times10^{9}$ cm; \citealt{Kallman09,M08}), then the total optical depth could easily approach 5, and the observed wind column density and the Compton scattering optical depth can plausibly be reconciled.

The second issue is that the Comptonized spectrum is sensitive to the (unknown) compactness of the scattering medium (e.g., \citealt{Coppi99}), so in order for the same {\tt eqpair} analysis of Section \ref{sec:corona} to hold, the change in the size of the scattering medium (at least two orders of magnitude from the corona to the wind) would have to be effectively balanced by changes in the seed photon luminosity and electron acceleration power (i.e., the wind heating/driving mechanism). For a much lower compactness (e.g., for a fixed luminosity and a much larger scattering region), Compton cooling is negligible compared to bremsstrahlung cooling and Coulomb collisions (\citealt{Coppi99}), which provides a physical motivation for Model 3.

\subsubsection{Bremsstrahlung from the wind}
\label{sec:wind}
Here we focus on the bremsstrahlung component in Model 3, where the change in the spectrum around MJD 53459 is explained not by a change in the temperature of some scattering medium, but by the fact that around this date, the cold bremsstrahlung component begins to make a significant contribution to the  observed X-ray flux. At the same time, as noted above, high-resolution X-ray spectra reveal an X-ray absorbing wind that is both dense and highly ionized -- precisely what is needed to produce significant bremsstrahlung emission. Here we ask: is the wind sufficiently hot and sufficiently luminous to produce the observed X-ray emission?

The bremsstrahlung normalization is given by 
\begin{equation}
K_{\rm br} = \frac{3.02\times10^{-15}}{4\pi D^2}\int n_e n_i dV,
\end{equation}
where $D=9.8\times10^{21}$ is the distance to the source in cm (\citealt{Orosz97}), $n_e$ is the electron density, $n_i$ is the ion density, and $V$ is the volume of the emitting region. Assuming a fully ionized spherical homogeneous emitter with ISM abundances (\citealt{Wilms00}), volume filling factor $f$, and radius $R$, we can simplify the normalization: $K_{\rm br}=8.5\times10^{-60}n_e^2R^3f.$ Plugging in the numbers from \citet{Kallman09} (quoted here in Section \ref{sec:coronawind}), we estimate $K_{\rm br,wind}=3f.$ 

Although this is below the observed normalization (typically in the range of 5-20 during the hypersoft state), it is important to remember that this $K_{\rm br,wind}=3f$ result is based on the photoionization calculations of \citet{M06a,M08,Kallman09}, which implicitly assume a point source geometry for the X-ray source. But we cannot take this geometry for granted, since both Models 2 and 3 are consistent with much or all of the observed X-ray emission emerging from the wind. On MJD 53470.5 (i.e., coinciding with our hypersoft state OIR SED), the bremsstrahlung normalization is $14_{-5}^{+7}$. If we set $f=1$ and integrate from $R=7\times10^{9}$ cm to $R=R_{\rm ex}=5.2\times10^{11}$ cm, we can reproduce $K_{\rm br}=14$ with an average density of $n_e=(3.1\pm0.6)\times10^{12}$ cm$^{-3},$ which is comparable to what is typically assumed for disk winds. If the density instead falls off like $R^{-2},$ we will have $K_{\rm br}\sim14$ for $n_e=(1.1\pm0.2)\times10^{15}$ cm$^{-3}$ at $R=7\times10^{9}$ cm. It is clear that the details are model dependent, but the result is unexpected: if we use mass continuity to extend a wind similar to the one described by \citet{Kallman09}, that wind could have a sufficiently large volume emission measure to contribute to the X-ray spectrum observed during the hypersoft state.

There are several important caveats for Model 3. First, bremsstrahlung emission should have associated atomic emission lines, which are not observed. Since \citet{Kallman09} report absorption line optical depths as high as 40, it is possible that the thermal emission lines are suppressed by the high column density absorber, but it would be unlikely to completely hide an emission component or a P-Cygni profile (although in their very high spectral resolution analysis, \citealt{Miller15} do find evidence for some such broadened emission lines from disk winds; see also \citealt{Miller16_arxiv}). Furthermore, we have only modeled the emission from optically thin gas, but for the density distributions derived from the bremsstrahlung normalizations in the preceding paragraph, the electron scattering optical depth can be well above 1 (see Section \ref{sec:compton}). This is not directly indicative of bremsstrahlung self-absorption, but it is evident that full radiative transfer would be needed to describe the emission in this model. Finally, given the high optical depths and the fact that optically thick thermal Comptonization provides a much better fit to the data, we prefer Model 2 to Model 3.

\subsubsection{The temperature of the wind}
\label{sec:temp}
We have established that an extended, dense wind could have a sufficiently large optical depth in electrons to produce the thermal Comptonization spectrum of the hypersoft state. But are the electrons in the wind hot enough to match the observed temperature of 2 keV? It is not entirely clear what the electron temperature in the wind should be. For instance, the Compton temperature of the hypersoft spectrum is just under 1 keV (see \citealt{Rahoui10,N12b} and references therein), but if the X-ray continuum is produced locally in the wind (whether by Compton scattering or bremsstrahlung), then the observed radiation is unlikely to be responsible for heating the gas. If either of these models is accurate, we have no constraint on the intrinsic radiation field during the hypersoft state: some or all of the inner disk is highly obscured.

For reference, the Compton temperature in the pre-hypersoft state X-ray observations reaches 7-8 keV (\citealt{N12b}). We conclude that 2 keV is a reasonable temperature for Compton-heated electrons during the hypersoft state. This is not to suggest that the wind must be driven by Compton heating, but simply to indicate that plausible estimates of the electron temperature in the wind bracket what is actually observed in the X-ray spectrum. If the wind is driven by magnetic processes (e.g., \citealt{M08}), the electron temperature could conceivably be rather different than in Compton heating scenarios. 

\subsubsection{A Compton-thick wind}
\label{sec:compton}

It appears that if we accept that the observed X-ray spectrum is well described by {\tt nthcomp} and that the absorption lines are produced in a narrow region of an extended, predominantly fully-ionized wind, then it is difficult to avoid the conclusion that the wind in its entirety (i.e., not just the X-ray absorber) is Compton thick. As noted above, the equivalent column density of the X-ray absorption line system in the wind is already $\sim 2/3$ of the Compton-thick limit, and any additional fully-ionized material will increase the total optical depth in the wind (see a similar suggestion from \citealt{Uttley15}). By way of example, in the uniform density scenario for Model 3 in Section \ref{sec:wind}, the electron optical depth is $\tau\sim1,$ and this rises to  $\tau\sim5.3$ in the $n_{\rm e}\propto R^{-2}$ case. These are rough estimates, since they are sensitive to the integration limits and the spectral model: Model 3 treats only a fraction of the emission as arising in the wind, leading to lower apparent bremsstrahlung normalizations and electron scattering optical depths than Model 2.

\citet{Reynolds12} details the conditions under which such a Compton-thick wind can and cannot be launched from the inner accretion disk by radiation pressure and magnetocentrifugal acceleration. These conditions amount to constraints on the ratio $\tau/\lambda$ as a function of radius, where $\lambda$ is the Eddington ratio of the inner accretion flow. Ideally we would apply that analysis to GRO J1655--40 to determine whether or not our explanation is plausible (see \citealt{N13a}). However, because our work indicates a substantial uncertainty in $\lambda,$ $\tau$, and the location of the wind, we shall not consider these constraints in detail.

It is worth noting, however, that our inference of a Compton-thick wind leads to a natural explanation for the non-detection of the inner accretion flow during the hypersoft state in Model 2: we simply cannot see it due to obscuration by the wind. The radiation is absorbed and scattered by the high column density wind. At lower inclination, the source could appear to be much more luminous, by a factor of over two orders of magnitude if our estimates of $\tau$ are applicable. We therefore echo the suggestion of \citet{Uttley15} that the intrinsic luminosity of the accretion flow could be at or above the Eddington limit (see also Section \ref{sec:bbody}). Because thermal driving can be boosted by radiation pressure at high Eddington ratios (see \citealt{PK02,N11a} and references therein), and because our best models require an alternative geometry for the X-ray emission and absorption, it may be illuminating to revisit the physical origin of this wind using models that are capable of unifying the X-ray and optical/infrared emission.

\subsection{The Blackbody Excess}
\label{sec:bbody}

\begin{figure}
\centerline{\includegraphics[width=2.25 in,angle=90,clip,trim=90 105 100 100]{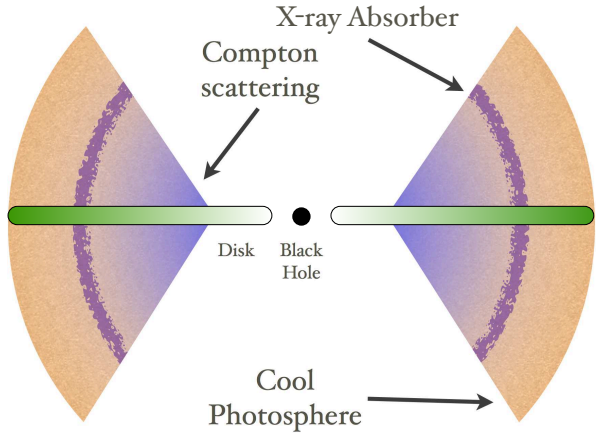}}
\caption{A cartoon of the geometry during the hypersoft state of \gro. The hot, dense interior of the wind produces Compton scattered emission that falls off rapidly with distance from the black hole. The X-ray absorbing portion of the wind lies in a narrow region in ionization and physical space exterior to the majority of the Compton scattered emission. The cooler outer photosphere emits blackbody radiation in the OIR. See text for details.\label{fig:fig}}
\end{figure}

If we suppose that an extended Compton-thick wind can explain the origin of the soft, curved X-ray continuum and the absence of standard disk/corona radiation during the hypersoft state, can it also explain the blackbody excess in the optical and infrared? Since the wind is optically thick, any radiation it absorbs should be thermalized and re-emitted in the outer regions. Indeed, \citet{King03} demonstrated that Compton-thick winds from black holes accreting near or above the Eddington limit should produce blackbody-like emission from an outer photosphere (see also \citealt{Begelman06,Poutanen07,Middleton15,Shen15,Soria16,Urquhart16}), where the optical depth is equal to unity. We present a cartoon of this scenario in Figure \ref{fig:fig}. Here we explore whether this model can explain a $\sim7,000$ K blackbody in GRO J1655--40. 

In an extension of the \citet{Shen15} model, \citet{Soria16} derive the radius and temperature of the blackbody photosphere $R_{\rm bb}$ and $T_{\rm bb}$ for a spherical geometry as
\begin{eqnarray}
R_{\rm bb} &=& 35.2~{\rm cm}\, \left( \frac{0.83\epsilon_{\rm w} - 0.25 \epsilon^2_{\rm w}}{f_{\rm v}} \right)^{20/11} \, 
    \left(1-\epsilon_{\rm w}\right)^{-21/22} \nonumber\\  
     &\times& (1+X)^{-27/22} \, m^{19/22} \, \dot{m}^{30/11} \left(1 + \frac{3}{5} \ln \dot{m} \right)^{-21/22} 
\end{eqnarray}
and
\begin{eqnarray}
T_{\rm bb} &=& 4.10 \times 10^9~\rm{K} \, \left( \frac{f_{\rm v}}{0.83\epsilon_{\rm w} - 0.25 \epsilon^2_{\rm w}} \right)^{10/11} \left(1-\epsilon_{\rm w}\right)^{8/11} \nonumber\\  
     &\times& (1+X)^{4/11} m^{-2/11} \dot{m}^{-15/11} \left(1 + \frac{3}{5} \ln \dot{m} \right)^{8/11},
\end{eqnarray}
where $\epsilon_{\rm w}$ is the fraction of the accretion power that powers the wind, $f_{\rm v}$ is the ratio of the wind speed to the escape speed evaluated at the launch radius of the wind, $X$ is the hydrogen mass fraction, $m$ is the black hole mass in $M_{\sun}$, and $\dot{m}$ is the external mass accretion rate in Eddington units. The accretion disk wind velocity during the 2005 outburst was measured at 375 \kms \citep{M08, Kallman09}, which corresponds to $f_{\rm v}\sim0.03-0.07$ for plausible launch radii (\citealt{N13a}). Here we assume solar abundances ($X=0.73$).

We find that we are able to reproduce the observed radius and temperature of the excess blackbody during the hypersoft state ($R_{\rm bb}=R_{\rm ex}=(5.2\pm0.2)\times10^{11}$ cm and $T_{\rm bb}=T_{\rm ex}=(7200\pm400)$ K, respectively) with $\epsilon_{\rm w}\sim0.9999$ and $\dot{m}=35-75.$ That is, if the available mass supply is highly super-Eddington, and essentially all of the available accretion power goes to accelerating an energy-driven wind, we expect the OIR excess observed in GRO J1655--40. For these parameters, using Equation 23 from \citet{Poutanen07}, the mass loss rate in the wind is between 17 and 40 times the Eddington rate. For comparison, the constant density spherical wind model described above would have a mass loss rate $>14\, \dot{M}_{\rm Edd}$ (c.f.\ \citealt{L02}). These numbers are very large but are quoted for a spherical geometry, and the actual values may be somewhat lower. \citet{Miller15} assume a covering factor of 0.2 (see also \citealt{M06a}). The model also predicts the radiative luminosity at the base of the wind (see also \citealt{SS73,Poutanen07}): $L/L_{\rm Edd}=(1+0.6\ln\dot{m})=3.1-3.6.$ In other words, the \gro\ accretion disk itself was likely super-Eddington during the hypersoft state if our analysis is correct (\citealt{Uttley15} come to the same conclusion based on timing analysis). 

There are, however, some reasons for caution when applying this model to GRO J1655--40. First, the model assumes that electron scattering dominates the opacity. For a fully ionized wind, this is sensible, but if our assertion is that the photosphere is 7,000 K, the absorption opacity may not be negligible in the outer regions of the wind. Indeed, the prescription for absorption opacity in \citet{Soria16} is comparable to electron scattering opacity for the high densities in the wind and for temperatures below $\sim10^{5}$ K, so it is plausible that this assumption is violated. This is important for understanding the radial ionization, density, and velocity profiles in the wind, but ultimately may require detailed radiative transfer models. This is especially necessary because the apparently very low ionization parameter of the outer wind (corresponding to $T\sim7,000$ K) is $\sim5$ orders of magnitude below that inferred for the X-ray absorber. Thus, depending on the opacity profile, the quantity $n_{e}R^2$ may need to increase significantly with radius, in which case the assumption of a constant velocity in the wind would also need revision. On a related note, it should be clear that our interpretation depends on the frequency-dependent opacity in this outflow: higher energy photons can emerge from deeper within the flow (\citealt{Shen15}; see also \citealt{Urquhart16}). 

We therefore tentatively conclude that the hypersoft state data are consistent with models of supercritical Compton thick winds. This conclusion must be tentative because we appear to be in a regime of parameter space where the assumptions of these models may not be entirely accurate, and may require a complete physical model. For the moment, we  leave this tension unresolved, but we note that these observations present an exciting challenge for models of winds, and we encourage future theoretical efforts to explain the data presented here.

If the OIR excess can indeed be related to the photosphere of a Compton-thick wind, we may finally be  in a position to consider the origin of the smooth modulation in the SMARTS light curves. A simple, plausible explanation for this periodicity would be that the photosphere of the wind is not completely axisymmetric (whether due to a warp in the disk, asymmetries in the driving mechanism, or some interaction with the accretion stream), such that the effective area of the photosphere is a function of orbital phase. The amplitude of the orbital oscillation is also similar to the typical amplitude of superhumps, as noted above, so it is possible that the modulation is due to the precession of an asymmetry in the disk excited by the outburst or the wind itself.

\section{Conclusion}
By all accounts, the 2005 outburst of GRO J1655--40 was remarkable. Although it began much like any other black hole outburst, significant departures from the norm are apparent in the first few weeks of observations. As detailed by \citet{Uttley15}, after a fairly typical spectrally hard state, instead of entering a normal spectrally soft state, the source transitioned to a `hypersoft' state, characterized by a steep hard X-ray spectrum ($\Gamma>5$ at 15--20 keV) that is not well described by the standard disk-plus-powerlaw model. Moreover, this hypersoft state coincided with (1) a period of unusual evolution in the timing properties of the system (specifically, a bend frequency that was extremely sensitive to the X-ray flux; \citealt{Uttley15}), (2) extremely deep absorption lines from an exceptional accretion disk wind (previously argued to be driven by magnetic processes; \citealt{M06a,Netzer06,DiazTrigo07,M08,Kallman09,N12b}) and, (3) as we have reported here, an unusual excess in the optical, infrared, and UV emission from the accretion flow. This excess is steady for the duration of the hypersoft state, although it appears to vary with binary orbital phase. Importantly, it appears to be much too large to be attributable to the accretion disk or the companion star.

It is our contention here that these ``coincidences" are not coincidences at all. The hot, dense accretion disk wind that is clearly observed during the hypersoft state can produce a steep, curved X-ray spectrum via Compton scattering in the low temperature/high optical depth regime, with a possible contribution from bremsstrahlung cooling. The physical variability of this wind, i.e., variations in its column density and optical depth, may be partly responsible for the unusual timing properties of the hypersoft state (see Model C in \citealt{Uttley15}), but our analysis indicates an additional possibility not considered by \citeauthor{Uttley15}: some of the intrinsic variability is itself produced in the wind. We suggest here that the very same extended wind could be responsible for the blackbody excess and its orbital variability. As we argue in Section \ref{sec:star}, efforts to explain the OIR data without reference to the wind are unsatisfactory. In contrast, however, models of energy driven super-Eddington Compton-thick winds (e.g., \citealt{Poutanen07,Shen15,Soria16}) can successfully reproduce the observed radius and temperature of the blackbody excess during the hypersoft state. Given the low temperature of our photosphere, we note that these models, which treat the absorption opacity as negligible relative to scattering, should be applied with caution, and we encourage additional theoretical work to confirm or refute our claim. 

It may not be necessary to impress upon the reader that we have proposed a fairly extreme scenario to explain these observations of GRO J1655--40: an accretion disk wind that is so dense and so extended that it not only obscures the bright inner accretion disk and produces its own X-ray emission but also has its own blackbody-emitting photosphere. Without drastically reducing the distance to the source and causing the secondary to fail to fill its Roche lobe, it is difficult to explain this OIR excess without reference to an extended object, and a Compton-thick wind can possibly explain such a photosphere and more. Although a robust estimate of the mass loss rate in the wind requires detailed simulations to account for its true energy-dependent opacity and geometry, estimates based on theoretical descriptions of Compton-thick outflows dominated by electron scattering opacity (\citealt{Soria16} and references therein) indicate that it could be $>17\times$ the Eddington rate (again, this may need to be reduced for a smaller covering factor). While this an extreme mass loss rate, the hypersoft state exhibits a set of unique multiwavelength timing and spectral behaviors, and it should not come as a surprise that an extreme scenario is required for a comprehensive explanation of the data. But whether or not our interpretation of the OIR emission is accurate, it is a robust conclusion of our work that the origin of the wind and the origin of the X-ray continuum cannot be understood independently, and that there is ample room for progress to be made in this area.

It should be noted, however, that the models applied here were not developed to explain GRO J1655--40. A number of quasars and ultraluminous X-ray sources (ULXs) have exhibited strong blackbody-like components in their X-ray spectra, and some (ultraluminous supersoft sources, ULSs) are dominated by this component. In recent years, it has become apparent that this emission and its variability can be produced by the photosphere of a Compton thick wind, launched by a super-Eddington accretion flow (e.g., \citealt{Mukai03,Fabbiano03,King03,Begelman06,Poutanen07,Shen15,Middleton15,Soria16,Urquhart16} and references therein). As clearly laid out by \citet{Begelman06} and \citet{Poutanen07}, these sources are highly anisotropic, which explains why the same model can be applied to ULXs and SS\,433. Our results here draw clear parallels between GRO J1655--40 and these high accretion rate objects, and it seems that the hypersoft state in the 2005 outburst of this black hole may have been a short-lived analog of the more persistent ULX behavior. This connection in fact makes the properties of the hypersoft state seems somewhat less extreme, as \citet{Urquhart16} find that ULSs may generally require $\dot{m}\sim$ a few 100, an order of magnitude higher than required here for GRO J1655--40.

One pressing question remains unanswered: why did this wind appear at all? Several lines of evidence indicate a very large mass accretion rate onto the black hole (producing radiation that is not visible from our line of sight due to obscuration by the wind). But there are to date no other confirmed examples of such an extreme outflow from a stellar-mass black hole transient (although \citealt{Soria00} found evidence of a transient opaque wind in their optical spectra of the 1994 outburst of \gro, and \citealt{U09} reported a similar high low temperature, high optical depth scatterer associated with many absorption lines in GRS 1915+105), and if our re-envisioned geometry is correct, there are no clear indications of the processes that triggered its formation. What caused the accretion rate to spike so sharply for $\sim50$ days? Is it even possible to achieve such high accretion rates in GRO J1655-40? Our estimates here clearly pose challenges to the standard picture of black hole outbursts, and this highlights the necessity of improving our treatment of the opacity in the wind for accurate estimates of $\dot{m}$. At present, we cannot address this using our data, but if indeed the OIR excess can be reliably associated with a super-Eddington Compton-thick wind, perhaps it can be used in the future to identify such winds in new transients, and perhaps these systems will shed additional light on the physics of winds around stellar mass black holes.

\acknowledgements We are extremely grateful to the referee for useful comments that significantly improved the paper. J.N.\ gratefully acknowledges funding support from NASA through the Hubble Postdoctoral Fellowship program, grant HST-HF2-51343.001-A, and through the Einstein Postdoctoral Fellowship program, grant PF2-130097. We thank John Raymond, Chris Done, Megumi Shidatsu, Emrah Kalemci, Roberto Soria, Jason Dexter, Chris Reynolds, Mike Nowak, Gabriele Ponti, Claude Canizares, and Alan Marscher for helpful discussions. We are also indebted to the International Space Science Institute, where some of this work was discussed, for their hospitality.

\bibliographystyle{apj_set3}
\bibliography{ms}

\begin{thebibliography}{120}
\expandafter\ifx\csname natexlab\endcsname\relax\def\natexlab#1{#1}\fi

\bibitem[{{Begelman} {et~al.}(2006){Begelman}, {King}, \&
  {Pringle}}]{Begelman06}
{Begelman}, M.~C., {King}, A.~R., \& {Pringle}, J.~E. 2006, \mnras, 370, 399

\bibitem[{{Begelman} {et~al.}(1983){Begelman}, {McKee}, \& {Shields}}]{B83}
{Begelman}, M.~C., {McKee}, C.~F., \& {Shields}, G.~A. 1983, \apj, 271, 70

\bibitem[{{Belloni}(2010{\natexlab{a}})}]{Belloni10b}
{Belloni}, ed. 2010{\natexlab{a}}, Lecture Notes in Physics, Berlin Springer
  Verlag, Vol. 794, {The Jet Paradigm}

\bibitem[{{Belloni}(2010{\natexlab{b}})}]{Belloni10a}
{Belloni}, T.~M. 2010{\natexlab{b}}, in Lecture Notes in Physics, Berlin
  Springer Verlag, Vol. 794, Lecture Notes in Physics, Berlin Springer Verlag,
  ed. {T.~Belloni}, 53--+

\bibitem[{{Blandford} \& {Payne}(1982)}]{BlandfordPayne}
{Blandford}, R.~D., \& {Payne}, D.~G. 1982, \mnras, 199, 883

\bibitem[{{Brocksopp} {et~al.}(2006){Brocksopp}, {McGowan}, {Krimm}, {Godet},
  {Roming}, {Mason}, {Gehrels}, {Still}, {Page}, {Moretti}, {Shrader},
  {Campana}, \& {Kennea}}]{Brocksopp06}
{Brocksopp}, C., {et~al.} 2006, \mnras, 365, 1203

\bibitem[{{Buxton} \& {Bailyn}(2005)}]{Buxton05}
{Buxton}, M., \& {Bailyn}, C. 2005, The Astronomer's Telegram, 485, 1

\bibitem[{{Buxton} {et~al.}(2005){Buxton}, {Bailyn}, \& {Maitra}}]{Buxton05a}
{Buxton}, M., {Bailyn}, C., \& {Maitra}, D. 2005, The Astronomer's Telegram,
  418, 1

\bibitem[{{Buxton} \& {Bailyn}(2004)}]{Buxton04}
{Buxton}, M.~M., \& {Bailyn}, C.~D. 2004, \apj, 615, 880

\bibitem[{{Canizares} {et~al.}(2005){Canizares}, {Davis}, {Dewey}, {Flanagan},
  {Galton}, {Huenemoerder}, {Ishibashi}, {Markert}, {Marshall}, {McGuirk},
  {Schattenburg}, {Schulz}, {Smith}, \& {Wise}}]{C05}
{Canizares}, C.~R., {et~al.} 2005, \pasp, 117, 1144

\bibitem[{{Coppi}(1999)}]{Coppi99}
{Coppi}, P.~S. 1999, in Astronomical Society of the Pacific Conference Series,
  Vol. 161, High Energy Processes in Accreting Black Holes, ed. J.~{Poutanen}
  \& R.~{Svensson}, 375

\bibitem[{{Corbel} {et~al.}(2013){Corbel}, {Coriat}, {Brocksopp}, {Tzioumis},
  {Fender}, {Tomsick}, {Buxton}, \& {Bailyn}}]{Corbel13}
{Corbel}, S., {et~al.} 2013, \mnras, 428, 2500

\bibitem[{{Corbel} \& {Fender}(2002)}]{Corbel02}
{Corbel}, S., \& {Fender}, R.~P. 2002, \apjl, 573, L35

\bibitem[{{Corbel} {et~al.}(2000){Corbel}, {Fender}, {Tzioumis}, {Nowak},
  {McIntyre}, {Durouchoux}, \& {Sood}}]{Corbel00}
{Corbel}, S., {et~al.} 2000, \aap, 359, 251

\bibitem[{{Corbel} {et~al.}(2003){Corbel}, {Nowak}, {Fender}, {Tzioumis}, \&
  {Markoff}}]{Corbel03}
---. 2003, \aap, 400, 1007

\bibitem[{{Coriat} {et~al.}(2009){Coriat}, {Corbel}, {Buxton}, {Bailyn},
  {Tomsick}, {K{\"o}rding}, \& {Kalemci}}]{Coriat09}
{Coriat}, M., {et~al.} 2009, \mnras, 400, 123

\bibitem[{{Coriat} {et~al.}(2011){Coriat}, {Corbel}, {Prat}, {Miller-Jones},
  {Cseh}, {Tzioumis}, {Brocksopp}, {Rodriguez}, {Fender}, \&
  {Sivakoff}}]{Coriat11}
---. 2011, \mnras, 414, 677

\bibitem[{{Davies}(1990)}]{Davies90}
{Davies}, S.~R. 1990, \mnras, 244, 93

\bibitem[{{DePoy} {et~al.}(2003){DePoy}, {Atwood}, {Belville}, {Brewer},
  {Byard}, {Gould}, {Mason}, {O'Brien}, {Pappalardo}, {Pogge}, {Steinbrecher},
  \& {Teiga}}]{Depoy03}
{DePoy}, D.~L., {et~al.} 2003, in Society of Photo-Optical Instrumentation
  Engineers (SPIE) Conference Series, Vol. 4841, Instrument Design and
  Performance for Optical/Infrared Ground-based Telescopes, ed. M.~{Iye} \&
  A.~F.~M. {Moorwood}, 827--838

\bibitem[{{D{\'{\i}}az Trigo} \& {Boirin}(2013)}]{DiazTrigo13a}
{D{\'{\i}}az Trigo}, M., \& {Boirin}, L. 2013, Acta Polytechnica, 53, 659

\bibitem[{{D{\'{\i}}az Trigo} {et~al.}(2014){D{\'{\i}}az Trigo}, {Migliari},
  {Miller-Jones}, \& {Guainazzi}}]{DiazTrigo14}
{D{\'{\i}}az Trigo}, M., {et~al.} 2014, \aap, 571, A76

\bibitem[{{D{\'{\i}}az Trigo} {et~al.}(2007){D{\'{\i}}az Trigo}, {Parmar},
  {Miller}, {Kuulkers}, \& {Caballero-Garc{\'{\i}}a}}]{DiazTrigo07}
---. 2007, \aap, 462, 657

\bibitem[{{Done} \& {Gierli{\'n}ski}(2003)}]{Done03}
{Done}, C., \& {Gierli{\'n}ski}, M. 2003, \mnras, 342, 1041

\bibitem[{{Done} {et~al.}(2007){Done}, {Gierli{\'n}ski}, \& {Kubota}}]{Done07}
{Done}, C., {Gierli{\'n}ski}, M., \& {Kubota}, A. 2007, \aapr, 15, 1

\bibitem[{{Ebisawa}(1997)}]{Ebisawa97a}
{Ebisawa}, K. 1997, in X-Ray Imaging and Spectroscopy of Cosmic Hot Plasmas,
  ed. {F.~Makino \& K.~Mitsuda}, 427--+

\bibitem[{{Eggleton}(1983)}]{Eggleton83}
{Eggleton}, P.~P. 1983, \apj, 268, 368

\bibitem[{{Esin} {et~al.}(1997){Esin}, {McClintock}, \& {Narayan}}]{E97}
{Esin}, A.~A., {McClintock}, J.~E., \& {Narayan}, R. 1997, \apj, 489, 865

\bibitem[{{Fabbiano} {et~al.}(2003){Fabbiano}, {King}, {Zezas}, {Ponman},
  {Rots}, \& {Schweizer}}]{Fabbiano03}
{Fabbiano}, G., {et~al.} 2003, \apj, 591, 843

\bibitem[{{Falcke} {et~al.}(2004){Falcke}, {K{\"o}rding}, \&
  {Markoff}}]{Falcke04}
{Falcke}, H., {K{\"o}rding}, E., \& {Markoff}, S. 2004, \aap, 414, 895

\bibitem[{{Fender} {et~al.}(1999){Fender}, {Corbel}, {Tzioumis}, {McIntyre},
  {Campbell-Wilson}, {Nowak}, {Sood}, {Hunstead}, {Harmon}, {Durouchoux}, \&
  {Heindl}}]{Fender99}
{Fender}, R., {et~al.} 1999, \apjl, 519, L165

\bibitem[{{Fender}(2001)}]{Fender01b}
{Fender}, R.~P. 2001, \mnras, 322, 31

\bibitem[{{Fender} {et~al.}(2004){Fender}, {Belloni}, \& {Gallo}}]{FBG04}
{Fender}, R.~P., {Belloni}, T.~M., \& {Gallo}, E. 2004, \mnras, 355, 1105

\bibitem[{{Fender} {et~al.}(2009){Fender}, {Homan}, \& {Belloni}}]{Fender09}
{Fender}, R.~P., {Homan}, J., \& {Belloni}, T.~M. 2009, \mnras, 396, 1370

\bibitem[{{Fitzpatrick}(1999)}]{Fitzpatrick99}
{Fitzpatrick}, E.~L. 1999, \pasp, 111, 63

\bibitem[{{Foellmi} {et~al.}(2006){Foellmi}, {Depagne}, {Dall}, \&
  {Mirabel}}]{Foellmi06}
{Foellmi}, C., {et~al.} 2006, \aap, 457, 249

\bibitem[{{Frank} {et~al.}(2002){Frank}, {King}, \& {Raine}}]{FKR02}
{Frank}, J., {King}, A., \& {Raine}, D.~J. 2002, {Accretion Power in
  Astrophysics: Third Edition} (Cambridge, UK: Cambridge University Press)

\bibitem[{{Gallo} {et~al.}(2006){Gallo}, {Fender}, {Miller-Jones}, {Merloni},
  {Jonker}, {Heinz}, {Maccarone}, \& {van der Klis}}]{Gallo06}
{Gallo}, E., {et~al.} 2006, \mnras, 370, 1351

\bibitem[{{Gallo} {et~al.}(2003){Gallo}, {Fender}, \& {Pooley}}]{Gallo03}
{Gallo}, E., {Fender}, R.~P., \& {Pooley}, G.~G. 2003, \mnras, 344, 60

\bibitem[{{Gallo} {et~al.}(2012){Gallo}, {Miller}, \& {Fender}}]{Gallo12}
{Gallo}, E., {Miller}, B.~P., \& {Fender}, R. 2012, \mnras, 423, 590

\bibitem[{{Gandhi} {et~al.}(2011){Gandhi}, {Blain}, {Russell}, {Casella},
  {Malzac}, {Corbel}, {D'Avanzo}, {Lewis}, {Markoff}, {Cadolle Bel}, {Goldoni},
  {Wachter}, {Khangulyan}, \& {Mainzer}}]{Gandhi11}
{Gandhi}, P., {et~al.} 2011, \apjl, 740, L13

\bibitem[{{Gehrels} {et~al.}(2004){Gehrels}, {Chincarini}, {Giommi}, {Mason},
  {Nousek}, {Wells}, {White}, {Barthelmy}, {Burrows}, {Cominsky}, {Hurley},
  {Marshall}, {M{\'e}sz{\'a}ros}, {Roming}, {Angelini}, {Barbier}, {Belloni},
  {Campana}, {Caraveo}, {Chester}, {Citterio}, {Cline}, {Cropper}, {Cummings},
  {Dean}, {Feigelson}, {Fenimore}, {Frail}, {Fruchter}, {Garmire}, {Gendreau},
  {Ghisellini}, {Greiner}, {Hill}, {Hunsberger}, {Krimm}, {Kulkarni}, {Kumar},
  {Lebrun}, {Lloyd-Ronning}, {Markwardt}, {Mattson}, {Mushotzky}, {Norris},
  {Osborne}, {Paczynski}, {Palmer}, {Park}, {Parsons}, {Paul}, {Rees},
  {Reynolds}, {Rhoads}, {Sasseen}, {Schaefer}, {Short}, {Smale}, {Smith},
  {Stella}, {Tagliaferri}, {Takahashi}, {Tashiro}, {Townsley}, {Tueller},
  {Turner}, {Vietri}, {Voges}, {Ward}, {Willingale}, {Zerbi}, \&
  {Zhang}}]{Gehrels04}
{Gehrels}, N., {et~al.} 2004, \apj, 611, 1005

\bibitem[{{Gierli{\'n}ski} {et~al.}(2009){Gierli{\'n}ski}, {Done}, \&
  {Page}}]{Gierlinski09}
{Gierli{\'n}ski}, M., {Done}, C., \& {Page}, K. 2009, \mnras, 392, 1106

\bibitem[{{Gierli{\'n}ski} {et~al.}(1999){Gierli{\'n}ski}, {Zdziarski},
  {Poutanen}, {Coppi}, {Ebisawa}, \& {Johnson}}]{Gierlinski99}
{Gierli{\'n}ski}, M., {et~al.} 1999, \mnras, 309, 496

\bibitem[{{Greene} {et~al.}(2001){Greene}, {Bailyn}, \& {Orosz}}]{Greene01}
{Greene}, J., {Bailyn}, C.~D., \& {Orosz}, J.~A. 2001, \apj, 554, 1290

\bibitem[{{G{\"u}ltekin} {et~al.}(2009){G{\"u}ltekin}, {Cackett}, {Miller}, {Di
  Matteo}, {Markoff}, \& {Richstone}}]{Gultekin09}
{G{\"u}ltekin}, K., {et~al.} 2009, \apj, 706, 404

\bibitem[{{Hannikainen} {et~al.}(1998){Hannikainen}, {Hunstead},
  {Campbell-Wilson}, \& {Sood}}]{Hannikainen98}
{Hannikainen}, D.~C., {et~al.} 1998, \aap, 337, 460

\bibitem[{{Hjellming} \& {Rupen}(1995)}]{Hjellming95}
{Hjellming}, R.~M., \& {Rupen}, M.~P. 1995, \nat, 375, 464

\bibitem[{{Homan} \& {Belloni}(2005)}]{Homan05b}
{Homan}, J., \& {Belloni}, T. 2005, \apss, 300, 107

\bibitem[{{Homan} {et~al.}(2005){Homan}, {Buxton}, {Markoff}, {Bailyn},
  {Nespoli}, \& {Belloni}}]{Homan05}
{Homan}, J., {et~al.} 2005, \apj, 624, 295

\bibitem[{{Houck}(2002)}]{Houck02}
{Houck}, J.~C. 2002, in High Resolution X-ray Spectroscopy with XMM-Newton and
  Chandra, ed. {G.~Branduardi-Raymont} (London: MSSL), 17

\bibitem[{{Houck} \& {Denicola}(2000)}]{HD00}
{Houck}, J.~C., \& {Denicola}, L.~A. 2000, in ASP Conf. Ser., Vol. 216,
  Astronomical Data Analysis Software and Systems IX, ed. {N.~Manset,
  C.~Veillet, \& D.~Crabtree} (San Francisco, CA: ASP), 591

\bibitem[{{Hynes} {et~al.}(2003){Hynes}, {Haswell}, {Cui}, {Shrader},
  {O'Brien}, {Chaty}, {Skillman}, {Patterson}, \& {Horne}}]{Hynes03}
{Hynes}, R.~I., {et~al.} 2003, \mnras, 345, 292

\bibitem[{{Hynes} {et~al.}(1998){Hynes}, {Haswell}, {Shrader}, {Chen}, {Horne},
  {Harlaftis}, {O'Brien}, {Hellier}, \& {Fender}}]{Hynes98}
---. 1998, \mnras, 300, 64

\bibitem[{{Kallman} {et~al.}(2009){Kallman}, {Bautista}, {Goriely}, {Mendoza},
  {Miller}, {Palmeri}, {Quinet}, \& {Raymond}}]{Kallman09}
{Kallman}, T.~R., {et~al.} 2009, \apj, 701, 865

\bibitem[{{King} {et~al.}(2013){King}, {Miller}, {Raymond}, {Fabian},
  {Reynolds}, {G{\"u}ltekin}, {Cackett}, {Allen}, {Proga}, \&
  {Kallman}}]{King13}
{King}, A.~L., {et~al.} 2013, \apj, 762, 103

\bibitem[{{King} {et~al.}(2012){King}, {Miller}, {Raymond}, {Fabian},
  {Reynolds}, {Kallman}, {Maitra}, {Cackett}, \& {Rupen}}]{King12}
---. 2012, \apjl, 746, L20

\bibitem[{{King} {et~al.}(2014){King}, {Walton}, {Miller}, {Barret}, {Boggs},
  {Christensen}, {Craig}, {Fabian}, {F{\"u}rst}, {Hailey}, {Harrison},
  {Krivonos}, {Mori}, {Natalucci}, {Stern}, {Tomsick}, \& {Zhang}}]{King14}
---. 2014, \apjl, 784, L2

\bibitem[{{King} \& {Pounds}(2003)}]{King03}
{King}, A.~R., \& {Pounds}, K.~A. 2003, \mnras, 345, 657

\bibitem[{{K{\"o}rding} {et~al.}(2006){K{\"o}rding}, {Fender}, \&
  {Migliari}}]{Kording06}
{K{\"o}rding}, E.~G., {Fender}, R.~P., \& {Migliari}, S. 2006, \mnras, 369,
  1451

\bibitem[{{Leahy} \& {Leahy}(2015)}]{Leahy15}
{Leahy}, D.~A., \& {Leahy}, J.~C. 2015, Computational Astrophysics and
  Cosmology, 2, 4

\bibitem[{{Lee} {et~al.}(2002){Lee}, {Reynolds}, {Remillard}, {Schulz},
  {Blackman}, \& {Fabian}}]{L02}
{Lee}, J.~C., {et~al.} 2002, \apj, 567, 1102

\bibitem[{{Luketic} {et~al.}(2010){Luketic}, {Proga}, {Kallman}, {Raymond}, \&
  {Miller}}]{Luketic10}
{Luketic}, S., {et~al.} 2010, \apj, 719, 515

\bibitem[{{Maccarone}(2003)}]{Maccarone03}
{Maccarone}, T.~J. 2003, \aap, 409, 697

\bibitem[{{Markoff} {et~al.}(2001){Markoff}, {Falcke}, \&
  {Fender}}]{Markoff01a}
{Markoff}, S., {Falcke}, H., \& {Fender}, R. 2001, \aap, 372, L25

\bibitem[{{Markoff} {et~al.}(2005){Markoff}, {Nowak}, \& {Wilms}}]{Markoff05}
{Markoff}, S., {Nowak}, M.~A., \& {Wilms}, J. 2005, \apj, 635, 1203

\bibitem[{{Marshall} {et~al.}(2006){Marshall}, {Robin}, {Reyl{\'e}},
  {Schultheis}, \& {Picaud}}]{Marshall06}
{Marshall}, D.~J., {et~al.} 2006, \aap, 453, 635

\bibitem[{{McClintock} \& {Remillard}(2006)}]{McClintockRemillard06}
{McClintock}, J.~E., \& {Remillard}, R.~A. 2006, {Black hole binaries}, ed.
  {Lewin, W.~H.~G.~\& van der Klis, M.}, Cambridge Astrophysics Series, No. 39
  (Cambridge, UK: Cambridge University Press), 157--213

\bibitem[{{Merloni} {et~al.}(2003){Merloni}, {Heinz}, \& {di
  Matteo}}]{Merloni03}
{Merloni}, A., {Heinz}, S., \& {di Matteo}, T. 2003, \mnras, 345, 1057

\bibitem[{{Middleton} {et~al.}(2015){Middleton}, {Heil}, {Pintore}, {Walton},
  \& {Roberts}}]{Middleton15}
{Middleton}, M.~J., {et~al.} 2015, \mnras, 447, 3243

\bibitem[{{Migliari} {et~al.}(2007){Migliari}, {Tomsick}, {Markoff}, {Kalemci},
  {Bailyn}, {Buxton}, {Corbel}, {Fender}, \& {Kaaret}}]{Migliari07}
{Migliari}, S., {et~al.} 2007, \apj, 670, 610

\bibitem[{{Miller} {et~al.}(2015){Miller}, {Fabian}, {Kaastra}, {Kallman},
  {King}, {Proga}, {Raymond}, \& {Reynolds}}]{Miller15}
{Miller}, J.~M., {et~al.} 2015, \apj, 814, 87

\bibitem[{{Miller} {et~al.}(2006){Miller}, {Raymond}, {Fabian}, {Steeghs},
  {Homan}, {Reynolds}, {van der Klis}, \& {Wijnands}}]{M06a}
---. 2006, \nat, 441, 953

\bibitem[{{Miller} {et~al.}(2016){Miller}, {Raymond}, {Fabian}, {Gallo},
  {Kaastra}, {Kallman}, {King}, {Proga}, {Reynolds}, \&
  {Zoghbi}}]{Miller16_arxiv}
---. 2016, ApJ, in press. arxiv:1603.04714

\bibitem[{{Miller} {et~al.}(2008){Miller}, {Raymond}, {Reynolds}, {Fabian},
  {Kallman}, \& {Homan}}]{M08}
---. 2008, \apj, 680, 1359

\bibitem[{{Mitsuda} {et~al.}(1984){Mitsuda}, {Inoue}, {Koyama}, {Makishima},
  {Matsuoka}, {Ogawara}, {Suzuki}, {Tanaka}, {Shibazaki}, \&
  {Hirano}}]{Mitsuda84}
{Mitsuda}, K., {et~al.} 1984, \pasj, 36, 741

\bibitem[{{Motta} {et~al.}(2012){Motta}, {Homan}, {Mu{\~n}oz Darias},
  {Casella}, {Belloni}, {Hiemstra}, \& {M{\'e}ndez}}]{Motta12}
{Motta}, S., {et~al.} 2012, \mnras, 427, 595

\bibitem[{{Mu{\~n}oz-Darias} {et~al.}(2011){Mu{\~n}oz-Darias}, {Motta}, \&
  {Belloni}}]{MunozDarias11}
{Mu{\~n}oz-Darias}, T., {Motta}, S., \& {Belloni}, T.~M. 2011, \mnras, 410, 679

\bibitem[{{Mukai} {et~al.}(2003){Mukai}, {Pence}, {Snowden}, \&
  {Kuntz}}]{Mukai03}
{Mukai}, K., {et~al.} 2003, \apj, 582, 184

\bibitem[{{Neilsen}(2013)}]{N13a}
{Neilsen}, J. 2013, Advances in Space Research, 52, 732

\bibitem[{{Neilsen} {et~al.}(2014){Neilsen}, {Coriat}, {Fender}, {Lee},
  {Ponti}, {Tzioumis}, {Edwards}, \& {Broderick}}]{N14a}
{Neilsen}, J., {et~al.} 2014, \apjl, 784, L5

\bibitem[{{Neilsen} \& {Homan}(2012)}]{N12b}
{Neilsen}, J., \& {Homan}, J. 2012, \apj, 750, 27

\bibitem[{{Neilsen} \& {Lee}(2009)}]{NL09}
{Neilsen}, J., \& {Lee}, J.~C. 2009, \nat, 458, 481

\bibitem[{{Neilsen} {et~al.}(2012){Neilsen}, {Petschek}, \& {Lee}}]{N12a}
{Neilsen}, J., {Petschek}, A.~J., \& {Lee}, J.~C. 2012, \mnras, 2287

\bibitem[{{Neilsen} {et~al.}(2011){Neilsen}, {Remillard}, \& {Lee}}]{N11a}
{Neilsen}, J., {Remillard}, R.~A., \& {Lee}, J.~C. 2011, \apj, 737, 69

\bibitem[{{Netzer}(2006)}]{Netzer06}
{Netzer}, H. 2006, \apjl, 652, L117

\bibitem[{{Nowak} {et~al.}(2012){Nowak}, {Wilms}, {Pottschmidt}, {Schulz},
  {Maitra}, \& {Miller}}]{Nowak12b}
{Nowak}, M.~A., {et~al.} 2012, \apj, 744, 107

\bibitem[{{Orosz} \& {Bailyn}(1997)}]{Orosz97}
{Orosz}, J.~A., \& {Bailyn}, C.~D. 1997, \apj, 477, 876

\bibitem[{{Ponti} {et~al.}(2012){Ponti}, {Fender}, {Begelman}, {Dunn},
  {Neilsen}, \& {Coriat}}]{Ponti12}
{Ponti}, G., {et~al.} 2012, \mnras, 422, L11

\bibitem[{{Poutanen} {et~al.}(2007){Poutanen}, {Lipunova}, {Fabrika},
  {Butkevich}, \& {Abolmasov}}]{Poutanen07}
{Poutanen}, J., {et~al.} 2007, \mnras, 377, 1187

\bibitem[{{Proga}(2000)}]{Proga2000}
{Proga}, D. 2000, \apj, 538, 684

\bibitem[{{Proga}(2003)}]{Proga03}
---. 2003, \apj, 585, 406

\bibitem[{{Proga} \& {Kallman}(2002)}]{PK02}
{Proga}, D., \& {Kallman}, T.~R. 2002, \apj, 565, 455

\bibitem[{{Rahoui} {et~al.}(2010){Rahoui}, {Chaty}, {Rodriguez}, {Fuchs},
  {Mirabel}, \& {Pooley}}]{Rahoui10}
{Rahoui}, F., {et~al.} 2010, \apj, 715, 1191

\bibitem[{{Rahoui} {et~al.}(2012){Rahoui}, {Coriat}, {Corbel}, {Cadolle Bel},
  {Tomsick}, {Lee}, {Rodriguez}, {Russell}, \& {Migliari}}]{Rahoui12}
---. 2012, \mnras, 422, 2202

\bibitem[{{Rahoui} {et~al.}(2011){Rahoui}, {Lee}, {Heinz}, {Hines},
  {Pottschmidt}, {Wilms}, \& {Grinberg}}]{Rahoui11}
---. 2011, \apj, 736, 63

\bibitem[{{Remillard} \& {McClintock}(2006)}]{RM06}
{Remillard}, R.~A., \& {McClintock}, J.~E. 2006, \araa, 44, 49

\bibitem[{{Reynolds}(2012)}]{Reynolds12}
{Reynolds}, C.~S. 2012, \apjl, 759, L15

\bibitem[{{Roming} {et~al.}(2005){Roming}, {Kennedy}, {Mason}, {Nousek}, {Ahr},
  {Bingham}, {Broos}, {Carter}, {Hancock}, {Huckle}, {Hunsberger}, {Kawakami},
  {Killough}, {Koch}, {McLelland}, {Smith}, {Smith}, {Soto}, {Boyd},
  {Breeveld}, {Holland}, {Ivanushkina}, {Pryzby}, {Still}, \&
  {Stock}}]{Roming05}
{Roming}, P.~W.~A., {et~al.} 2005, \ssr, 120, 95

\bibitem[{{Russell} {et~al.}(2006){Russell}, {Fender}, {Hynes}, {Brocksopp},
  {Homan}, {Jonker}, \& {Buxton}}]{Russell06}
{Russell}, D.~M., {et~al.} 2006, \mnras, 371, 1334

\bibitem[{{Russell} {et~al.}(2007){Russell}, {Maccarone}, {K{\"o}rding}, \&
  {Homan}}]{Russell07}
---. 2007, \mnras, 379, 1401

\bibitem[{{Russell} {et~al.}(2010){Russell}, {Maitra}, {Dunn}, \&
  {Markoff}}]{Russell10}
---. 2010, \mnras, 405, 1759

\bibitem[{{Shakura} \& {Sunyaev}(1973)}]{SS73}
{Shakura}, N.~I., \& {Sunyaev}, R.~A. 1973, \aap, 24, 337

\bibitem[{{Shen} {et~al.}(2015){Shen}, {Barniol Duran}, {Nakar}, \&
  {Piran}}]{Shen15}
{Shen}, R.-F., {et~al.} 2015, \mnras, 447, L60

\bibitem[{{Smak}(2006)}]{Smak06}
{Smak}, J.~I. 2006, \actaa, 56, 365

\bibitem[{{Soria} \& {Kong}(2016)}]{Soria16}
{Soria}, R., \& {Kong}, A. 2016, \mnras, 456, 1837

\bibitem[{{Soria} {et~al.}(2000){Soria}, {Wu}, \& {Hunstead}}]{Soria00}
{Soria}, R., {Wu}, K., \& {Hunstead}, R.~W. 2000, \apj, 539, 445

\bibitem[{{Steiner} {et~al.}(2009){Steiner}, {Narayan}, {McClintock}, \&
  {Ebisawa}}]{Steiner09b}
{Steiner}, J.~F., {et~al.} 2009, \pasp, 121, 1279

\bibitem[{{Sunyaev} \& {Titarchuk}(1980)}]{Sunyaev80}
{Sunyaev}, R.~A., \& {Titarchuk}, L.~G. 1980, \aap, 86, 121

\bibitem[{{Tananbaum} {et~al.}(1972){Tananbaum}, {Gursky}, {Kellogg},
  {Giacconi}, \& {Jones}}]{Tananbaum72}
{Tananbaum}, H., {et~al.} 1972, \apjl, 177, L5+

\bibitem[{{Ueda} {et~al.}(2009){Ueda}, {Yamaoka}, \& {Remillard}}]{U09}
{Ueda}, Y., {Yamaoka}, K., \& {Remillard}, R. 2009, \apj, 695, 888

\bibitem[{{Urquhart} \& {Soria}(2016)}]{Urquhart16}
{Urquhart}, R., \& {Soria}, R. 2016, \mnras, 456, 1859

\bibitem[{{Uttley} \& {Klein-Wolt}(2015)}]{Uttley15}
{Uttley}, P., \& {Klein-Wolt}, M. 2015, \mnras, 451, 475

\bibitem[{{van der Hooft} {et~al.}(1998){van der Hooft}, {Heemskerk},
  {Alberts}, \& {van Paradijs}}]{vanderHooft98}
{van der Hooft}, F., {et~al.} 1998, \aap, 329, 538

\bibitem[{{Veledina} {et~al.}(2013){Veledina}, {Poutanen}, \&
  {Vurm}}]{Veledina13}
{Veledina}, A., {Poutanen}, J., \& {Vurm}, I. 2013, \mnras, 430, 3196

\bibitem[{{Wegner}(2003)}]{Wegner03}
{Wegner}, W. 2003, Astronomische Nachrichten, 324, 219

\bibitem[{{Wilms} {et~al.}(2000){Wilms}, {Allen}, \& {McCray}}]{Wilms00}
{Wilms}, J., {Allen}, A., \& {McCray}, R. 2000, \apj, 542, 914

\bibitem[{{Woods} {et~al.}(1996){Woods}, {Klein}, {Castor}, {McKee}, \&
  {Bell}}]{Woods96}
{Woods}, D.~T., {et~al.} 1996, \apj, 461, 767

\bibitem[{{Zdziarski} {et~al.}(1996){Zdziarski}, {Johnson}, \&
  {Magdziarz}}]{Zdziarski96}
{Zdziarski}, A.~A., {Johnson}, W.~N., \& {Magdziarz}, P. 1996, \mnras, 283, 193

\bibitem[{{Zdziarski} {et~al.}(2010){Zdziarski}, {Misra}, \&
  {Gierli{\'n}ski}}]{Zdziarski10}
{Zdziarski}, A.~A., {Misra}, R., \& {Gierli{\'n}ski}, M. 2010, \mnras, 402, 767

\bibitem[{{{\.Z}ycki} {et~al.}(1999){{\.Z}ycki}, {Done}, \& {Smith}}]{Zycki99}
{{\.Z}ycki}, P.~T., {Done}, C., \& {Smith}, D.~A. 1999, \mnras, 309, 561

\end{thebibliography}
\label{lastpage}

\end{document}